\documentclass[12pt]{iopart}

\usepackage{iopams}  
\usepackage{perpage}
\usepackage{graphicx, epsfig} 
\usepackage{color}
\usepackage{url}
\usepackage{epstopdf}

\def\nn{\nonumber}

\newcommand{\E}{e}
\newcommand{\beq}{\begin{eqnarray}}
\newcommand{\eeq}{\end{eqnarray}}

\newcommand{\ba}{\begin{eqnarray}}
\newcommand{\ea}{\end{eqnarray}}

\newcommand{\be}{\begin{equation}}
\newcommand{\ee}{\end{equation}}

\def\L5{\tilde{\Lambda}}

\begin{document}

\title[Black holes in massive gravity]{Black holes in massive gravity}

\author{Eugeny Babichev$^1$ and Richard Brito$^2$}

\address{$^1$ Laboratoire de Physique Th\'eorique d'Orsay,
B\^atiment 210, Universit\'e Paris-Sud 11,
F-91405 Orsay Cedex, France.}

\address{$^2$ CENTRA, Departamento de F\'{\i}sica, Instituto Superior T\'ecnico, Universidade de Lisboa,
Avenida Rovisco Pais 1, 1049 Lisboa, Portugal.}

\eads{\mailto{eugeny.babichev@th.u-psud.fr}, \mailto{richard.brito@tecnico.ulisboa.pt}} 

\begin{abstract}
We review the black hole solutions of the ghost-free massive gravity theory and its bimetric extension and outline the main results on the stability of these solutions against small perturbations. 
Massive (bi)-gravity accommodates exact black hole solutions, analogous to those of General Relativity. In addition to these solutions, hairy black holes -- solutions with no correspondent in General Relativity -- have been found numerically, 
whose existence is a natural consequence of the absence of Birkhoff's theorem in these theories.
The existence of extra propagating degrees of freedom, makes
the stability properties of these black holes richer and more complex than those of General Relativity.
In particular, the bi-Schwarzschild black hole exhibits an unstable spherically symmetric mode, while the bi-Kerr geometry is also generically unstable, both against the spherical mode and against superradiant instabilities. If astrophysical black holes are described by these solutions, the superradiant instability of the Kerr solution imposes stringent bounds on the graviton mass.
\end{abstract}

\maketitle

\section{Introduction}

Massive gravity is a modification of General Relativity (GR) based on the idea of equipping the graviton with mass. 
A model of non self-interacting massive gravitons was first suggested by Fierz and Pauli in the early beginnings of field theory~\cite{Pauli:1939xp,Fierz:1939ix}. This original model was subsequently shown by van Dam, Veltman and Zakharov (vDVZ)  to differ from GR even at small distance scales, 
ruling out the theory on the basis of Solar system tests~\cite{vanDam:1970vg}.
A solution to this problem was later conjectured by Vainshtein~\cite{Vainshtein:1972sx}, 
who argued that GR could be recovered at small distances by including non-linear terms in the field equations of the hypothetical massive gravity theory. Much later, rigorous studies of several non-linear completions of massive gravity showed that this would indeed generically be the case (for a review see Ref.~\cite{Babichev:2013usa}).
However, generic nonlinear versions of the Fierz-Pauli theory, although able to recover GR via the Vainshtein mechanism, 
turned out to reveal another pathology --- the so-called Boulware-Deser ghost~\cite{Boulware:1973my}. 
This ghost problem has only been solved recently in a series of works~\cite{deRham:2010kj,deRham:2010ik,deRham:2011rn,Hassan:2011tf,Hassan:2012qv,Hassan:2011hr,Hassan:2011ea}, 
where it was shown that for a subclass of  massive potentials
the Boulware-Deser ghost does not appear, both in massive gravity --- a theory with one dynamical and one fixed metric, the so-called de Rham, Gabadadze and Tolley (dRGT) model --- and 
its bi-gravity extension (see Refs.~\cite{Hinterbichler:2011tt,deRham:2014zqa} for recent reviews on massive gravity)\footnote{A settled term often used in the literature to refer to dRGT 
 theory and its extensions is the ``ghost-free massive gravity''. This name should be used with care, since dRGT theory is free from the Boulware-Deser ghost (or Ostrogradski ghost), 
 but the other degrees of freedom are not necessary healthy for particular solutions. Therefore the right term should rather be ``Boulware-Deser ghost free massive gravity'' or ``Ostrogradski ghost free massive gravity'', although we will also sometimes refer to dRGT theory and its extension simply as ``ghost free massive gravity'', keeping in mind the above reservation.}.
The bi-gravity theory contains two dynamical metrics, which interact with each other via non-derivative terms. 
If ordinary matter only couples to one of the metrics, then one can interpret the theory as an extension of Einstein's gravity with an extra spin-2 field coupled to 
gravity in a particular non-minimal way.

Despite its very recent formulation, a considerable amount of work has been done to understand the implications of the ghost-free theory. Although a great deal of the interest has been focused on building cosmological models out of these theories~(see~\cite{deRham:2014zqa} and references therein), an increasing effort is underway to construct and understand the properties of black hole (BH) solutions in massive (bi)-gravity. Understanding BHs in any theory of gravity is obviously of extreme importance, not only to understand the highly non-linear regime of the theory, but also to possibly look for deviations from GR (see e.g. Ref.~\cite{Berti:2015itd}). 

It is interesting that the first BH solutions in a nonlinear theory of massive gravity were found in the context of high-energy physics~\cite{Salam:1976as}, with no mention on the possible application of the theory as a modification of gravity. 
Reference~\cite{Salam:1976as} introduced non-bidiagonal solutions of a bi-metric theory, 
with both metrics having the Schwarzschild-(anti)de-Sitter form, but not diagonal simultaneously.
However, BHs in bi-gravity did not attract much attention until the discovery of the massive (bi)-gravity theory safe from the Boulware-Deser ghost. 
The discovery of this theory led to a plethora of works on the subject, and similar solutions have been put forward for both dRGT massive gravity
and its bi-gravity extension~\cite{Koyama:2011xz,Comelli:2011wq,Volkov:2012wp,Nieuwenhuizen:2011sq,Berezhiani:2011mt}. More interestingly, it has been shown that spherically symmetric solutions which do not exist in GR are also present~\cite{Volkov:2012wp,Brito:2013xaa}. These are the only examples known so far of ``hairy''  solutions in theories of massive gravity (see Refs.~\cite{VolkovTasinato,Volkov:2014ooa} for reviews covering this topic).
Later, Ref.~\cite{Babichev:2014fka} generalized some of these previous works finding a class of non-bidiagonal Reissner-Nordstr\"{o}m solutions in dRGT 
massive (bi)-gravity, while Ref.~\cite{Babichev:2014tfa} was able to construct a family of rotating BHs in the same theories.

Along with the search for BH solutions in massive (bi)-gravity, 
in the last couple of years progress has been made to understand the stability properties of these solutions. 
One of the most striking results of this ongoing effort was the conclusion that the bidiagonal Schwarzschild solution is unstable, 
as found in~\cite{Babichev:2013una} and confirmed independently in~\cite{Brito:2013wya}.
In contrast, the same type of instability has been shown to be absent for non-bidiagonal solutions~\cite{Babichev:2014oua}.

As a by-product of studying the stability of these solutions, by understanding how small fluctuations behave in theories with massive gravitons it is also possible to understand how gravitational waveforms might differ from GR. The extra gravitational polarizations and the nontrivial dispersion introduced by a putative small graviton mass may leave important imprints on gravitational waveforms from, e.g., inspiralling compact objects. Understanding these effects is thus necessary given that advanced gravitational-wave detectors~\cite{LIGO,VIRGO} will begin operation (Advanced LIGO~\cite{LIGO} is in fact expected to begin to collect data in mid-2015) and the first direct detection of gravitational waves is expected to take place within the next decade. With this is mind, Refs.~\cite{Brito:2013wya,Brito:2013yxa} studied generic perturbations of Schwarzschild and Kerr BHs in massive (bi)-gravity, with the particularly interesting result that the Kerr solution would be prone to another kind of instability, related to the superradiant scattering of bosonic fields with spinning BHs (see Refs.~\cite{Brito:2015oca,Cardoso:2013krh} for recent reviews).
This instability, which was known to occur for massive scalar~\cite{Detweiler:1980uk,Dolan:2007mj} and vector~\cite{Pani:2012bp,Pani:2012vp,Witek:2012tr} perturbations of rotating BHs in GR, was shown to occur in massive gravity due to the natural existence of linear massive spin-2 perturbations in the theory. If BHs in the Universe were to be described by these solutions, the most striking consequence of this instability is the existence of a bound in the graviton mass~\cite{Brito:2013wya}. 
 
This work is meant to review in some detail BH solutions in dRGT massive gravity and its bi-gravity extension, 
including both exact and numerical solutions, 
with the focus on the stability issues --- a fresh and fast developing topic.
We introduce in Sec.~\ref{Genres}  massive gravity theories: Fierz-Pauli theory, non-linear completions and the dRGT model with its bi-gravity extension. 
Then, in Sec.~\ref{BHSOLUTIONS} we present recent developments on BH solutions in dRGT model: the exact bidiagonal and non-bidiagonal solutions,
as well as numerical solutions. 
Sec.~\ref{PERTURBATIONS} is devoted to results of the past two years on perturbations and stability of BHs in massive (bi)-gravity.
In Sec.~\ref{CONCLUSIONS} we conclude with open issues and work in progress.

\section{Massive gravity}

 \label{Genres}

\subsection{Fierz-Pauli theory}
\label{PFsec}
Before introducing the theory of massive (bi)-gravity, we briefly review the linear massive gravity model, also known as the Fierz-Pauli theory~\cite{Pauli:1939xp,Fierz:1939ix}.
The linear theory can also be viewed as an expansion of the full non-linear massive gravity model around a Minkowski background. 
Expanding the Einstein-Hilbert action in metric perturbations $h_{\mu\nu}$
as $g_{\mu\nu}=\eta_{\mu\nu}+h_{\mu\nu}$, 
where $\eta_{\mu\nu}$ is the Minkowski metric $\eta_{\mu\nu}= \rm{diag}(-1,+1,+1,+1)$ and 
the indices of $h_{\mu \nu}$ are moved up and down with the metric $\eta_{\mu \nu}$, and keeping only quadratic terms in the action we obtain the linear approximation of GR,
\begin{equation}\label{GRq}
	S_{GR} = M_P^2\int d^4 x \sqrt{-g}R = M_P^2 \int d^{4}x
\left( -\frac12 h^{\mu\nu}\mathcal{E}^{\alpha\beta}_{\mu\nu} h_{\alpha\beta}\right) +\mathcal{O} (h^3),
\end{equation}
where $M_P$ is the Planck mass and 
\begin{equation*}
\fl
{\cal E}_{\mu \nu}  \equiv \mathcal{E}^{\alpha\beta}_{\mu\nu}h_{\alpha\beta} 
	= -\frac{1}{2} \partial_\mu \partial_\nu h - \frac{1}{2} \Box h_{\mu \nu} + \frac{1}{2} \partial_\rho \partial_\mu h^\rho_\nu 
		+ \frac{1}{2}\partial_\rho \partial_\nu h^\rho_\mu  - \frac{1}{2} \eta_{\mu \nu}(\partial^\rho \partial^\sigma h_{\rho \sigma} - \Box h),
\end{equation*}
is the linearized Einstein tensor 
$G_{\mu\nu} = {\cal E}_{\mu \nu}  +\mathcal{O} (h^3) $. 
When matter is present, the metric perturbation $h_{\mu\nu}$ is also coupled to the energy-momentum tensor $T_{\mu\nu}$, via the interaction term $h_{\mu\nu}T^{\mu\nu}$, 
but since here we will mostly consider vacuum solutions, we omit this term. 
The action (\ref{GRq}) contains only derivative terms. 
By adding non-derivative quadratic terms $h^2$ (where $h=h_{\mu \nu}\eta^{\mu \nu}$) and $h_{\mu\nu}h^{\mu\nu}$ to the action (\ref{GRq}), 
a linear massive gravity theory is obtained. If the non-derivative terms are added in a special combination 
$\propto \left( h_{\mu \nu}h^{\mu \nu}-h^2\right) $, the action takes the  Fierz-Pauli form~\cite{Pauli:1939xp,Fierz:1939ix}, 
\begin{equation}\label{actionPF}
S_{PF} = M_P^2 \int d^4x   \left[  -\frac12 h^{\mu\nu}\mathcal{E}^{\alpha\beta}_{\mu\nu} h_{\alpha\beta}
- \frac{1}{4} m^2 \left( h_{\mu\nu}h^{\mu\nu}-h^2\right) \right],
\end{equation}
where $m$ is the mass parameter, corresponding to the graviton mass.
This particular mass combination describes  the only consistent linear Lorentz-invariant theory for a massive spin-2 field.
Models with other mass terms can be shown to necessarily contain a physical ghost degree of freedom~\cite{Pauli:1939xp,Fierz:1939ix}, which is in turn related to the Ostrogradski ghost~\cite{Ostrogradski}.
The Fierz-Pauli theory (\ref{actionPF}) fails to pass the Solar system tests of gravity, due to the fact that 
a massive graviton has more polarizations than a massles one, therefore modifying the gravitational interaction.
Even in the zero graviton mass limit this leads to a difference between the massless and massive gravitation interaction which is known as the vDVZ discontinuity~\cite{vanDam:1970vg}.

\subsection{Non-linear massive gravity}

Starting from the quadratic action (\ref{actionPF}), one can try to guess a non-linear generalization of the theory, 
in the same way that full GR is a non-linear generalization of the quadratic action~(\ref{GRq}).
Obviously, the first term in~(\ref{actionPF}) constructed out of the metric $g$ should be the Einstein-Hilbert term\footnote{In principle, since Lovelock's theorem~\cite{Lovelock:1971yv} is not valid in massive gravity, other kinetic terms could be possible. However it was shown in Refs.~\cite{deRham:2013tfa} that any new kinetic term would generically lead to ghost instabilities.}. 
However, it is not possible to get a non-linear massive term, using only the metric $g_{\mu\nu}$, 
since the only nontrivial term corresponds to a Lagrangian density proportional $\sqrt{-g}$, which stands for the cosmological constant. 
Thus, one way to have a non-trivial mass term is to add a second metric, say $f_{\mu\nu}$, which can be chosen to be fixed 
(in this case the theory has a preferred background, i.e. ``aether'') or dynamical, in which case the theory is called bi-gravity or massive bimetric gravity. 
The metric $g_{\mu\nu}$ can be non-derivatively coupled to the second metric $f_{\mu\nu}$, in order to form a non-trivial mass term. 
Non-linear mass terms should be chosen such that: 
the action is invariant under a coordinate change common to both metrics;
there is a (almost) flat solution for $g_{\mu \nu}$; in the limit where $g_{\mu \nu}=\eta_{\mu\nu}+h_{\mu\nu}$ and  $f_{\mu\nu}=\eta_{\mu\nu}$
the potential at quadratic order for $h_{\mu \nu}$ takes the Fierz-Pauli form~(\ref{actionPF}). In spite of the restrictions formulated above, there is a huge freedom in choosing the mass term. In fact one can choose the interaction term in a class of functions satisfying these conditions (see e.g. Ref.~\cite{Damour:2002ws}). 
The term,  
$$
\sqrt{-g}\;(g_{\mu \nu} - f_{\mu \nu}) (g_{\sigma \tau} - f_{\sigma \tau}) \left(g^{\mu\sigma}g^{\nu\tau}-g^{\mu\nu}g^{\sigma\tau}\right),
$$
considered in \cite{Arkani-Hamed:2002sp}, is an example of such an interaction. 
As it has been proposed by Vainshtein~\cite{Vainshtein:1972sx}, 
the inclusion of non-linear terms in the equations of motion for a massive graviton help to solve the problem of the vDVZ discontinuity. 
However, as was later noticed in Ref.~\cite{Boulware:1973my} a proper proof of this conjecture was lacking. 
The status of the Vainshtein mechanism in non-linear massive gravity was in fact only solved recently, when analytic and numerical studies confirmed that the Vainshtein mechanism indeed works in this model~\cite{Babichev:2009us} (for a recent review see~\cite{Babichev:2013usa}).
Furthermore in the work by Boulware and Deser~\cite{Boulware:1973my} another serious problem has been found:
the non-linear interaction terms generically lead to ghost instabilities (appearing at the non-linear level). 
The Boulware-Deser instability can be thought as another face of the Ostrogradski ghost instability~\cite{Ostrogradski}. 

\subsection{dRGT gravity}
\label{dRGT}

Although,  in general,  the Boulware-Deser ghost persists in non-linear massive and bi-gravity theories, 
it has been found by de Rham, Gabadadze and Tolley (dRGT)  that 
in the so-called ``decoupling 
limit'' --- a limit where the degrees of freedom of the theory (almost) decouple --- there is a restricted subclass of potential terms, for which the Boulware-Deser ghost is absent even at the non-linear 
level~\cite{deRham:2010kj,deRham:2010ik,deRham:2011rn,Hassan:2012qv}. 
Later a full Hamiltonian analysis confirmed the absence of the Boulware-Deser ghost in this model~\cite{Hassan:2011tf,Hassan:2011hr,Hassan:2011ea}, while fully covariant proofs were given in Ref.~\cite{Deffayet:2012nr} for a subset of possible massive terms and for generic mass terms in~\cite{Kugo:2014hja} (see also the review~\cite{deRham:2014zqa}).

To formulate the dRGT theory and its extension to the bi-gravity case, 
it is convenient to introduce functions $\E_k$ of matrices $X$, 
which represent the elementary symmetric polynomials of the eigenvalues of $X$. 
In the case of $4\times 4$ matrices they are given by (cf. e.g.  \cite{Hassan:2011hr}),
\begin{equation}\label{eX}
\fl 
\eqalign{ 
e_0  &= 1,\;
e_1  = [X] ,\;
e_2 = \frac{1}{2} \left([X]^2 - [X^2]\right),\; e_3  = \frac{1}{6} \left([X]^3 - 3 [X][X^2] + 2 [X^3]\right), \\
e_4  &= \frac{1}{24} \left([X]^4 - 6[X]^2[X^2] +3[X^2]^2 + 8 [X][X^3]- 6 [X^4]\right),}
\end{equation}
where $[X]$ stands for the trace of $X$. Note that $e_4$ can also be written in the simpler way, $e_4 = {\rm det} (X)$.
The building block of the dRGT mass term is the square root of the matrix ${\bf g}^{-1}{\bf f}$, i.e. one defines the matrix $\gamma$ as follows,
\begin{equation}
\bf{\gamma} = \sqrt{{\bf g}^{-1}{\bf f}},
\end{equation}
with the matrix elements of $\bf{\gamma}$ defined as $\gamma^{\mu}_{\nu}=\sqrt{g^{\mu\lambda}f_{\lambda\nu}}$.
The action of the most general bi-gravity theory without the Boulware-Deser ghost then reads~\cite{Hassan:2011zd},
\be \label{dGTBETA}\fl
S = M_P^2 \int d^4 x \sqrt{-g}\left[R[g] - 2 m^2 \sum_{k=0}^{k=4} \beta_k \E_k\left(\gamma\right)\right] 
	+\kappa M^2_P\int d^4 x \sqrt{-f}\, \mathcal{R}[f] \,,
\ee
where $R[g]$ and $\mathcal{R}[f] $ are the Ricci scalars for the metrics $g$ and $f$ respectively,
$\beta_n$ and $\kappa$ are arbitrary coefficients. 
The bi-gravity action (\ref{dGTBETA}) contains both the kinetic term for the metric $g$ and one for the metric $f$.
The coefficient $\kappa$ marks the difference between the Planck masses for the $g$ and $f$ metrics.
The original dRGT theory~\cite{deRham:2010kj,deRham:2010ik,deRham:2011rn} (dRGT massive gravity) 
is given by the action~(\ref{dGTBETA}) dropping the last term, such that there is no dynamics for the metric $f$, contrary to the bi-gravity theory.
Note also that the terms $\beta_0$ and $\beta_4$ do not give a mass to the graviton. Since $\sqrt{-g}\; \E_4\left(\bf{\gamma}\right) = \sqrt{-g} \;{\rm det} \left(\bf{\gamma}\right) = \sqrt{-f}$, the $\beta_0$ term describes a cosmological constant for the metric $g$, while the $\beta_4$ term corresponds to the cosmological constant of the metric $f$. 
Thus, there is a three parameter family of massive bi-gravity theories parametrized by  $\beta_k$, with  $k=1,2,3$ (which becomes a two parameter family once the mass of the graviton is fixed). 

The action (\ref{dGTBETA}) can be written in an alternative form, using the matrix $\mathcal{K}$, defined as,
\begin{equation}
	\mathcal{K} = {\mathbb I} - \bf{\gamma}.
\end{equation}
The bidiagonal extension of the dRGT action then reads,
\begin{equation}
\eqalign{ 
\label{dGTBETAa}
S_{biG} = & M^2_P\int d^4 x \sqrt{-g} \left(R[g] +2 m^2 \mathcal{U}[g,f]  + 2 m^2 \Lambda_g \right) \\ 
&+ \kappa M^2_P\int d^4 x \sqrt{-f} \left(\mathcal{R}[f] + 2m^2 \Lambda_f \right), }
\end{equation}
where 
\begin{equation}\label{potential}
	\mathcal{U}[g,f] \equiv e_2(K) + \alpha_3 e_3(K)  + \alpha_4 e_4(K) ,
\end{equation}
with the following identifications for the coefficients, 
\begin{eqnarray}
\beta_0 = - \left(  \Lambda_g + 6 + 4\alpha_3 + \alpha_4 \right),\, \beta_1 = \left(3 + 3\alpha_3+ \alpha_4\right),\nonumber\\
 \beta_2 = -\left(1 +2\alpha_3+\alpha_4\right) ,\, \beta_3 = (\alpha_3+\alpha_4),\, \beta_4 = - \left( \kappa \Lambda_f + \alpha_4\right)\,,\nonumber
\end{eqnarray}
and where in the action~(\ref{dGTBETAa}) we explicitly separated the cosmological terms. 

The equations of motion derived from (\ref{dGTBETAa}) by variation with respect to $g_{\mu\nu}$ and $f_{\mu\nu}$ read,
\begin{eqnarray}
	G^{\mu}_{\phantom{\mu}\nu}  =&  m^2\left( T^{\mu}_{\phantom{\mu}\nu} -  \Lambda_g \delta^\mu_\nu \right),\label{Eg}\\
	\mathcal{G}^{\mu}_{\phantom{\mu}\nu}  = & m^2 \left(
		 \frac{\sqrt{-g}}{\sqrt{-f}}\frac{ \mathcal{T}^{\mu}_{\phantom{\mu}\nu}}{\kappa} - \Lambda_f \delta^\mu_\nu \right),\label{Ef}
\end{eqnarray}
where $G^{\mu}_{\phantom{\mu}\nu}$ and $\mathcal{G}^{\mu}_{\phantom{\mu}\nu}$ are the Einstein tensors for the metrics 
$g$ and $f$ correspondingly and the energy-momentum tensors coming from the interaction terms have the form
\begin{eqnarray}\label{Tmnexpl}
T_{\mu\nu} &= \mathcal{U} g_{\mu\nu} - 2 \frac{\delta \mathcal{U}}{\delta g^{\mu\nu}} = \nonumber\\
			& - g_{\mu\sigma}\gamma^{\sigma}_{\alpha}\left(\mathcal{K}^\alpha_\nu -[\mathcal{K}]\delta^\alpha_\nu\right)
			+\alpha_3 g_{\mu\sigma}\gamma^{\sigma}_{\alpha}\left(\mathcal{U}_2\delta^\alpha_\nu - [\mathcal{K}]\mathcal{K}^\alpha_\nu+(\mathcal{K}^2)^\alpha_\nu\right)\\
			&+\alpha_4 g_{\mu\sigma}\gamma^{\sigma}_{\alpha}\left(\mathcal{U}_3\delta^\alpha_\nu-\mathcal{U}_2\mathcal{K}^\alpha_\nu + [\mathcal{K}](\mathcal{K}^2)^\alpha_\nu-(\mathcal{K}^3)^\alpha_\nu\right) +\mathcal{U} g_{\mu\nu},\nonumber
\end{eqnarray}
\begin{eqnarray}\label{Tmnexplf}
\mathcal{T}_{\mu\nu} & = - 2 \frac{\delta \mathcal{U}}{\delta f^{\mu\nu}} =\nonumber\\
			&  f_{\mu\sigma}\gamma^{\sigma}_{\alpha}\left(\mathcal{K}^\alpha_\nu -[\mathcal{K}]\delta^\alpha_\nu\right)
			-\alpha_3 f_{\mu\sigma}\gamma^{\sigma}_{\alpha}\left(\mathcal{U}_2\delta^\alpha_\nu - [\mathcal{K}]\mathcal{K}^\alpha_\nu+(\mathcal{K}^2)^\alpha_\nu\right)\\
			&-\alpha_4 f_{\mu\sigma}\gamma^{\sigma}_{\alpha}\left(\mathcal{U}_3\delta^\alpha_\nu-\mathcal{U}_2\mathcal{K}^\alpha_\nu + [\mathcal{K}](\mathcal{K}^2)^\alpha_\nu-(\mathcal{K}^3)^\alpha_\nu\right).\nonumber
\end{eqnarray}
It can be shown that $T_{\mu\nu}$ and $\mathcal{T}_{\mu\nu}$ are symmetric~\cite{Baccetti:2012re}.
Note also a useful relation between the energy-momentum tensors,
\begin{equation}\label{Trel}
\mathcal{T}^\mu_{\phantom{\mu}\nu} = - T^\mu_{\phantom{\mu}\nu} + \mathcal{U}\delta^\mu_\nu,
\end{equation}
where $T^\mu_{\phantom{\mu}\nu}$ is found from $T_{\mu\nu}$ by raising an index with $g^{\mu\nu}$, 
while to get $\mathcal{T}^\mu_{\phantom{\mu}\nu}$ one raises an index with the metric $f^{\mu\nu}$. 
Furthermore the Bianchi identity implies the conservation conditions
\be
\nabla_g^{\mu}T_{\mu\nu}=0\,,\quad \nabla_f^{\mu}\mathcal{T}_{\mu\nu}=0\,, \label{bianchi1}\\
\ee
where $\nabla_g$ and $\nabla_f$ are the covariant derivatives with respect to $g_{\mu\nu}$ and $f_{\mu\nu}$ respectively. Note that due to Eq.~(\ref{Trel}) these two conditions are not independent. Finally, if one consider the metric $f_{\mu\nu}$ to be non-dynamical, then Eq.~(\ref{Ef}) must be excluded. 

\section{Black holes in massive (bi)-gravity}
\label{BHSOLUTIONS}
The structure of solutions in massive (bi)-gravity is more complex that in GR, mainly due to the fact that this theory has two metrics (see e.g. Ref.~\cite{Blas:2005yk} to see how the global structure of these solutions is affected by the co-existence of two metrics). In particular, the well-known Birkhoff's theorem for spherically symmetric solutions does not apply.
This suggests that in massive (bi)-gravity the classes of BH solutions are richer than in GR. 

Indeed, the spherically symmetric BH solutions in bi-gravity theories can be divided into two classes. 
The first class corresponds to the case for which the metrics cannot be brought simultaneously to a bidiagonal form. Said differently, in this class, 
if one metric in some coordinates is diagonal, the other metric is not.
BHs of the second class have two metrics which can be both written in the diagonal form, but not necessarily equal. 

The first BH solutions for a nonlinear massive gravity theory (suffering from the Boulware-Deser ghost) were constructed in Ref.~\cite{Salam:1976as}. 
Much later, spherically symmetric solutions for other classes of ghosty massive bi-gravity theories were found and classified  in detail in Refs.~\cite{Blas:2005yk,Berezhiani:2008nr}.
In the framework of the original dRGT model a class of non-bidiagonal Schwarzschild-de-Sitter solutions was found in Ref.~\cite{Koyama:2011xz}.
In Refs.~\cite{Comelli:2011wq,Volkov:2012wp}, 
spherically symmetric BH solutions were found for the bi-metric extension of dRGT theory, while 
spherically symmetric (charged and uncharged) solutions in the dRGT model for a special choice of the parameters of the action were presented in Refs.~\cite{Nieuwenhuizen:2011sq,Berezhiani:2011mt}. 
More recently, Ref.~\cite{Babichev:2014fka} found a more general class of charged BH solutions (in both the dRGT model and its bi-gravity extension). Finally Ref.~\cite{Babichev:2014tfa} generalized these findings by including rotation in the geometry. This last class of solutions, jointly with the charged solutions of Ref.~\cite{Babichev:2014fka},  includes as particular cases most of the previously found spherically symmetric solutions\footnote{With the exception of Schwarzschild non-bidiagonal solutions, presented in~\cite{Koyama:2011xz}, where an extra constant of integration
appears in the solution. However, in~\cite{Blas:2005yk} it has been argued (for similar BH solutions in a ghostly massive gravity) that the extra parameter should be set to a specific value for the solutions to be physical. In this case the solutions of~\cite{Koyama:2011xz} are a particular subclass of the solutions found in~\cite{Babichev:2014fka,Babichev:2014tfa}.
In~\cite{Volkov:2014ooa}, a method was presented to construct more general spherically symmetric non-bidiagonal solutions. These solutions are implicitly written in terms of one function (of two coordinates), which must satisfy a particular PDE, and thus describe an infinite-dimensional family of solutions (a similar technique has been used in~\cite{Mazuet:2015pea} to find de Sitter solutions in dRGT massive gravity). }.

Interestingly, spherically symmetric BH solutions with hair --- solutions differing from the Schwarzschild family --- were also found in bi-gravity theory, both with Anti-de Sitter~\cite{Volkov:2012wp} and flat asymptotics~\cite{Brito:2013xaa}.

Some good reviews on solutions of BHs in massive gravity already exist, e.g., Refs.~\cite{VolkovTasinato,Volkov:2014ooa}. Here we will mainly focus in giving a different and unified treatment of all the solutions found so far.

\subsection{Analytic solutions in massive (bi)-gravity}
\label{ana_sol}

Due to the complexity of the field equations, it turns out that it is easier to find analytical BH solutions of the first class (with non-bidiagonal metrics). 
In this section we mostly focus on this type of BHs, although some bidiagonal solutions will also be presented. Since the BH solutions of the original dRGT model can 
easily be obtained from the bi-metric ones, all our calculations will be done for the bi-metric theory described by the action~(\ref{dGTBETAa}).

To warm up, let us demonstrate the simplest BH solutions. 
We notice that for $g_{\mu\nu} = f_{\mu\nu}$ the potential term in the action~(\ref{dGTBETAa}) is zero, $\mathcal{U}[g,f] =0$. Comparison with GR then guarantee that in vacuum this theory admits the Kerr-(Anti) de Sitter metric as a solution. 
Taking for simplicity $\Lambda_g=\Lambda_f=0$, and considering the particular case of static solutions one then finds that the two metrics with the same Schwarzschild form,
\begin{eqnarray}
ds_g^2=ds_f^2  =  - \left(1-\frac{r_g}{r} \right) dt^2 +\frac{dr^2}{1-\frac{r_g}{r} }+r^2 d\Omega^2, \label{biSdS}
\end{eqnarray}
is a solution of the bi-gravity theory (\ref{dGTBETAa}).

\subsubsection{Static spherically symmetric solutions.}

To find other solutions it proves to be much more convenient to use coordinates which are regular at the horizon. One reason is that 
using such coordinates, we automatically avoid problems related to the singular behavior of the Schwarzschild coordinates at the horizon. In fact, for bidiagonal solutions this becomes a physical singularity, except for the particular case where the two metrics share the same horizon, like in the solution given in Eq.~(\ref{biSdS}).
More on this subject can be found in Ref.~\cite{Deffayet:2011rh} (see also Ref.~\cite{Banados:2011hk} where it was shown that the temperature of each horizon must also be the same). 

We start therefore with the advanced bi-Eddington-Finkelstein coordinates and assume the following ansatz for the two metrics~\cite{Babichev:2013una,Babichev:2014oua}, 
\begin{eqnarray}
ds_g^2 & = & -\left(1-\frac{r_g}{r}  -\frac{r^2}{l_g^2} \right) dv^2 +2dvdr+r^2 d\Omega^2,\label{gEF}\\
ds_f^2 & = & C^2\left[- \left(1-\frac{r_f}{r}  -\frac{r^2}{l_f^2} \right) dv^2 +2dvdr+r^2 d\Omega^2\right], \label{fEF}
\end{eqnarray}
where $r_g$ and $r_f$ are the Schwarzschild radii for the $g$-and $f$-metrics correspondingly, 
$l_g$ and $l_f$ are the ``cosmological'' radii and $C$ is a constant.  The Einstein tensors have the same form as in GR,
\begin{equation}\label{G}
	G^\mu_{\phantom{\mu}\nu} =
	-\frac{3}{l_g^2}\,\delta^\mu_\nu, \quad
	\mathcal{G}^\mu_{\phantom{\mu}\nu}  =  -\frac{3}{C^2 l_f^2}\,\delta^\mu_\nu,
\end{equation}
while the energy-momentum tensor (\ref{Tmnexpl}) takes the simple form,
\begin{equation}\label{Tmn}
T^{\mu}_{\phantom{\mu}\nu} =
\left(
\begin{array}{cccc}
  \lambda_g & 0 & 0 & 0 \\
 T^r_{\phantom{r}v} &  \lambda_g & 0 & 0 \\
 0 & 0 &  \lambda_g & 0 \\
 0 & 0 & 0 &   \lambda_g
\end{array}
\right),
\end{equation}
where
\begin{eqnarray}\label{Lg}
	\lambda_g =-(C-1)\left(\beta (C-1)^2-3 \alpha  (C-1)+3\right)\,,
\end{eqnarray}
is the effective cosmological constant for the metric $g$, 
\begin{eqnarray}\label{Toff}
	T^r_{\phantom{r}v}= - \frac{C}2 \left(\beta  (C-1)^2-2 \alpha  (C-1)+1\right) \left( \frac{r_g-r_f}{r}  +\frac{r^2}{l_g^2}-\frac{r^2}{l_f^2} \right)\,,
\end{eqnarray}
is the only non-diagonal term and we introduced the following constants,
\begin{equation}\label{ab}
\alpha\equiv 1+\alpha_3,\ \beta \equiv \alpha_3+\alpha_4\,.
\end{equation}
The energy-momentum tensor for the metric $f$ can then easily be found from Eqs.~(\ref{Tmn}) and~(\ref{Trel}). 
Clearly it must have the same form as Eq.~(\ref{Tmn}), but with different values for its components. Indeed, the diagonal part of 
$\mathcal{T}^\mu_{\phantom{\mu}\nu}$ consists of the components with values,
\begin{eqnarray}\label{Lf}
	 \lambda_f = C (C-1)\left( (C-1)^2 (1-\alpha+\beta) + 3(C-1)(1-\alpha)+ 3\right),
\end{eqnarray}
while the off-diagonal term reads,
\begin{eqnarray}\label{Tfoff}
	\mathcal{T}^r_{\phantom{r}v} = - T^r_{\phantom{r}v}\,.
\end{eqnarray}

To satisfy these modified Einstein's equations with the effective energy-momentum tensor given in Eq.~(\ref{Tmn}), the non-bidiagonal terms $\mathcal{T}^r_{\phantom{r}v}$ and  $T^r_{\phantom{r}v}$ must vanish. Depending on the way we ``kill'' those terms, different branches of solutions exist.
We either need to put to zero the expression in the first or in the second parentheses of Eq.~(\ref{Toff}).
First of all, one can easily see that the solution (\ref{biSdS}) is recovered for $r_g=r_f$, $l_g=l_f\to\infty$ and $C=1$ (in different coordinates though), 
since in this case $\mathcal{T}^\mu_{\phantom{\mu}\nu} =  T^\mu_{\phantom{\mu}\nu}  =0$.
A slightly more general solution is the bi-Schwarzschild-de-Sitter solution,
 which is obtained for $r_g=r_f$, $l_g=l_f$, $C=1$ and 
$\Lambda_g = \Lambda_f = 3/(m\, l_g)^2$.

Another class of bidiagonal solutions arises for $r_g=r_f$, $l_g=l_f$, but $C\neq 1$. 
In this case, the non-diagonal components of the energy-momentum tensors are zero, 
but the diagonal components $\lambda_g$ and $\lambda_f$ are not, so from Eqs.~(\ref{Eg}),~(\ref{Ef}) we have the following relations, 
\begin{eqnarray}\label{cnd2}
	3 = (m\, l_g)^2 \left( \Lambda_g - \lambda_g \right),\,\, 3 = C^2 (m\, l_g)^2 \left( \Lambda_f - \frac{\lambda_f}{C^4 \kappa} \right).
\end{eqnarray}
For a given set of parameters in the action~(\ref{dGTBETAa}), these equations determine $l_g$ and $C$ entering the metrics~(\ref{gEF}) and~(\ref{fEF}).

The non-bidiagonal class of BH solutions emerges for $r_f\neq r_g$ (and generically $l_f\neq l_g$).
In order to ``kill'' the non-bidiagonal terms $\mathcal{T}^r_{\phantom{r}v}$ and  $T^r_{\phantom{r}v}$
we need to satisfy the relation, 
\begin{equation}\label{cndab}
  \beta  (C-1)^2-2 \alpha  (C-1)+1 = 0,
\end{equation}
which fixes $C$ in terms of the parameters of the action, while
the de-Sitter radii are found from the field equations~(\ref{Eg}) and~(\ref{Ef}), which reduce to 
\begin{eqnarray}\label{cndnb}
	3 = (m\, l_g)^2 \left( \Lambda_g - \lambda_g \right),\,\, 3 = C^2 (m\, l_f)^2 \left( \Lambda_f - \frac{\lambda_f}{C^4 \kappa} \right).
\end{eqnarray}

\subsubsection{Black hole solutions with electric charge.} 

Adding an electromagnetic source to the theory given by action~(\ref{dGTBETAa}), it is also possible to find charged BH solutions.
We consider now the following action, 
\begin{equation}\label{actionM}
	S  = S_{biG}  -\frac14 \int d^4 x \sqrt{-g} F_{\mu\nu}F^{\mu\nu},
\end{equation}
where $S_{biG}$ is given by (\ref{dGTBETAa}) and the electromagnetic field is coupled to the metric $g$ only. 
The field equation~(\ref{Eg}) for the $g$-metric is then modified to include the energy-momentum tensor of the electromagnetic field,
\begin{eqnarray}
	G^{\mu}_{\phantom{\mu}\nu}  &=&  m^2\left( T^{\mu}_{\phantom{\mu}\nu} -  \Lambda_g \delta^\mu_\nu \right) 
		+ \frac{1}{M_P^2} \left( F^{\mu\alpha} F_{\nu\alpha}-\frac14\delta^\mu_\nu F^2 \right)\,,\label{EEMg}
\end{eqnarray}
while the field equation~(\ref{Ef}) for the metric $f$ remains the same.
We change accordingly the ansatz for the metric $g$ as follows~\cite{Babichev:2014fka},
\begin{eqnarray}
ds_g^2  =  -\left(1-\frac{r_g}{r} +\frac{r_Q^2}{r^2} -\frac{r^2}{l_g^2} \right) dv^2 +2dvdr+r^2 d\Omega^2. \label{gEMEF}
\end{eqnarray}
while keeping the ansatz for the metric $f$~(\ref{fEF}). 

With this modification the off-diagonal component of the energy-momentum tensor~(\ref{Toff}) is given by,
\begin{eqnarray}\label{TEMoff}
	T^r_{\phantom{r}v}= - \frac{C}2 \left(\beta  (C-1)^2-2 \alpha  (C-1)+1\right) \left( \frac{r_g-r_f}{r} -\frac{r_Q^2}{r^2} +\frac{r^2}{l_g^2}-\frac{r^2}{l_f^2} \right)\,.
\end{eqnarray}
Notice that the condition~(\ref{cndab}) still gives $T^r_{\phantom{r}v} = 0$ as before, 
while the condition~(\ref{cndnb}) provides the balance of the bare and the effective cosmological constants, fixing $l_g$ and $l_f$.
Taking the vector potential in the standard form,
\begin{equation}\label{A}
	A_\mu = \left\{ Q/r,0,0,0\right\},
\end{equation}
where $Q$ is the charge of the $g$-BH, we find that the field equations are satisfied if the conditions~(\ref{cndab}), (\ref{cndnb}) and 
\begin{equation}\label{Q}
	\sqrt{2}M_P r_Q  = Q\,,
\end{equation}
are also satisfied.
To summarise, a class of charged BH solutions for the theory~(\ref{actionM}) is given by the metrics~(\ref{gEMEF}) and~(\ref{fEF}), with the vector potential~(\ref{A}) and 
the conditions~(\ref{cndab}),~(\ref{cndnb}) and~(\ref{Q}).

\subsubsection{Enhanced symmetry of black hole solutions.} 
It turns out that for the particular combination of parameters of the action, $\beta = \alpha^2$,
the space of solutions is wider than in the generic case  $\beta \neq \alpha^2$, due to an enhanced symmetry for this choice of parameter.
Taking the following general ansatz for the metrics,
\begin{eqnarray}
ds_g^2 &=&  g_{vv} dv^2 +2 g_{vr} dvdr +g_{rr} dr^2 +r^2 d\Omega^2,\label{ges} \\
ds_f^2 & = & C^2\left[f_{vv} dv^2 +2 f_{vr} dvdr +f_{rr} dr^2 +r^2 d\Omega^2\right], \label{fes}
\end{eqnarray}
with $g_{vv}$, $g_{vr}$, $g_{rr}$, $f_{vv}$, $f_{vr}$, $f_{rr}$ functions of  $v$ and $r$, with the conditions $\beta=\alpha^2$ and (\ref{cndab}),
the energy-momentum tensors take the form of effective cosmological constants.
Note that in the metrics~(\ref{ges}) and (\ref{fes}) we only required spherical symmetry and not the precise ansatz for the metrics, 
in contrast to the Schwarzschild-de-Sitter ansatz (\ref{gEF}) and (\ref{fEF}) in the study above. 
Of course, we still need to satisfy the modified Einstein's equations with effective cosmological constants, 
thus the metrics $g$ and $f$ must have the Schwarzschild-de-Sitter form (or Reissner-Nordstr\"om-de Sitter, in case of non-zero charge) in some coordinates, 
as long as the conditions (\ref{cndnb}) are fulfilled. 
However, in this case, one of the metrics is not fixed with respect to the other, there is freedom to make a coordinate change of the form,
\begin{equation}\label{change}
v\to v(v,r),\, r\to r(v,r),
\end{equation}
in one metric, without touching the other metric at the same time.
One can perform an {\it independent} coordinate change (\ref{change}) for each metric and the result is also a solution. 
Therefore, the solution for $\beta=\alpha^2$ has an extra symmetry, which is absent in the general case.
In the case with a fixed metric $f$, the fact that there is an extra freedom in this solution has been discussed in detail in Ref.~\cite{Kodama:2013rea}.

In fact, several solutions for BHs presented in the literature fall in this general class of solutions.
In particular, the solutions of Ref.~\cite{Berezhiani:2011mt} and Ref.~\cite{Nieuwenhuizen:2011sq}, 
are examples of the solution described above, with particular coordinate changes of the form (\ref{change}) for one of the metrics $g$ and $f$, or both. 
More details can be found in Ref.~\cite{Babichev:2014fka}.

\subsubsection{Rotating black hole solutions}

The approach to look for BH solutions using the regular at the horizon metric ansatz 
also works --- with some modifications --- for rotating BHs~\cite{Babichev:2014tfa}.
Indeed, for the theory~(\ref{dGTBETAa}) let us assume the metrics $g$ and $f$ to have the form of 
the original Kerr metric element,
\begin{eqnarray}
\label{grotate}
\fl	ds_g^2  =  -\left( 1- \frac{r_g r}{\rho^2}\right)\left( dv+a\sin^2\theta d\phi \right)^2 \nonumber\\
			 + 2\left( dv+a\sin^2\theta d\phi \right)\left( dr+a\sin^2\theta d\phi \right) 
			+ \rho^2 \left( d\theta^2 +\sin^2\theta d\phi^2 \right).\\
\label{frotate}	
\fl ds_f^2 = C^2\left[
	 -\left( 1- \frac{2 r_f r}{\rho^2}\right)\left( dv+a\sin^2\theta d\phi \right)^2 \right.\nonumber\\
	\left.		 + 2\left( dv+a\sin^2\theta d\phi \right)\left( dr+a\sin^2\theta d\phi \right) 
			+ \rho^2 \left( d\theta^2 +\sin^2\theta d\phi^2 \right)\right].
\end{eqnarray}
where
\begin{equation}
	\rho^2 = r^2+ a^2\cos^2\theta,
\end{equation}
and $a$ is the rotation parameter. 
The ansatz~(\ref{grotate}) and (\ref{frotate}) can be viewed as an extension of~(\ref{gEF}) and (\ref{fEF}) for rotating BHs.
First of all, we note that since both metrics (\ref{grotate}) and (\ref{frotate}) coincide with the Kerr solution, then
\begin{equation}
	G_{\mu\nu} =  \mathcal{G}_{\mu\nu} =0.
\end{equation}
In our ansatz (\ref{grotate}) and (\ref{frotate}) the space for both metrics is asymptotically flat, 
in contrast to the previous cases (\ref{gEF}), (\ref{fEF}) and (\ref{gEMEF}), where we allowed asymptotically (A)dS space-times.
We have to set the right hand sides of (\ref{Eg}) and (\ref{Ef}) to zero in order to satisfy Einstein's equations, although one could straightforwardly generalize these solutions to the asymptotically (A)dS case. 

Lengthy but straightforward calculation of the energy-momentum tensors gives
\begin{equation}\label{TKerr}
 ^{(R)}T^{\mu}_{\phantom{\mu}\nu} =
\left(
\begin{array}{cccc}
  \lambda_{g} & 0 & 0 & 0 \\
^{(R)}T^r_{\phantom{r}v} &    \lambda_{g} & 0 &  ^{(R)}T^r_{\phantom{r}\phi } \\
 0 & 0 &   \lambda_{g} & 0 \\
 0 & 0 & 0 &     \lambda_{g}
\end{array}
\right),
\end{equation}
where $\lambda_g$  is given by (\ref{Lg}) and 
\begin{eqnarray}
	^{(R)}T^r_{\phantom{r}v} & = - \frac{C}2\left(\beta  (C-1)^2-2 \alpha  (C-1)+1\right)  \frac{(r_g-r_f) r}{\rho^2},\label{TKrv}\\
	^{(R)}T^r_{\phantom{r}\phi} & = - \frac{C}2 \left(\beta (C-1)^2-2 \alpha  (C-1)+1\right) \frac{a (r_g-r_f) r \sin^2\theta}{\rho^2}.\label{TKrp}
\end{eqnarray}
Here we use the superscript $(R)$ to emphasize that this corresponds to a rotating solution. Note that (\ref{TKrv}) coincides with (\ref{Toff}) for $a=0$ and $l_g=l_f=\infty$, while the r.h.s. of (\ref{TKrp}) is zero when there is no rotation, $a=0$.
From Eq.~(\ref{Trel}) it is now easy to find $^{(R)}\mathcal{T}^\mu_{\phantom{\mu}\nu}$: the off-diagonal components are given by,
\begin{eqnarray}\label{TfKoff}
	^{(R)}\mathcal{T}^r_{\phantom{r}v} = - ^{(R)}T^r_{\phantom{r}v}, \,\, ^{(R)}\mathcal{T}^r_{\phantom{r}\phi} = - ^{(R)}T^r_{\phantom{r}\phi},
\end{eqnarray}
while the diagonal components are equal to $\lambda_f$, given by Eq.~(\ref{Lf}).

Both off-diagonal components of (\ref{TKerr}) are zero if the condition~(\ref{cndab}) is satisfied (as in the spherically symmetric case).
On the other hand, the diagonal components, acting as an effective cosmological constant, must be balanced by 
an appropriately chosen $\Lambda_g$ and $\Lambda_f$,
\begin{eqnarray}\label{cndK}
	  \Lambda_g = \lambda_g ,\,\,  \Lambda_f = \frac{\lambda_f}{C^4 \kappa},
\end{eqnarray}
which is simply the old condition~(\ref{cnd2}) in the limit of asymptotically flat spacetimes, $l_g\to \infty$ and $l_f\to \infty$.

We have just shown that the metrics (\ref{grotate}) and (\ref{frotate}) with $C$ given by (\ref{cndab}) and $\Lambda_g$, $\Lambda_f$ 
given by (\ref{cndK}) are rotating BH solutions of the massive gravity model~(\ref{dGTBETAa}). 

It is not difficult now to get BH solutions of the dRGT model (with one fixed metric) 
from the BH solutions in bi-metric massive gravity.
The dRGT action is given by (\ref{dGTBETAa}) without the last term, which is the dynamical part of the second metric.
Therefore, one simply needs to retract the equations of motion of the second metric, 
assuming each time that the second metric is flat~\footnote{One can also consider the non-dynamical metric to be non-flat, where each choice corresponds to a different theory. In the particular case where the non-dynamical metric is flat, bidiagonal BH solutions do not exist, since these are necessarily singular at the horizon~\cite{Deffayet:2011rh}.}. 
In particular, in the case of the Schwarzschild-de-Sitter BH, 
the solution of dRGT model is given by (\ref{gEF}), and (\ref{fEF}) with $r_f=0$ and $l_f=\infty$, 
(so that $f_{\mu\nu}$ is flat although written in the ingoing Eddington-Finkelstein coordinates), 
with the condition (\ref{cndab}) and the first condition of (\ref{cndnb}). 
Similarly, a charged BH of the dRGT model is given by (\ref{gEMEF}), (\ref{fEF}) with $r_f=0$ and $l_f=\infty$,
the condition (\ref{cndab}), and the first condition of (\ref{cndnb}) and, additionally, by (\ref{Q}).
Finally, a rotating solution in dRGT is obtained from the solution of the bi-metric extension in a similar manner. 
In this case the metric $f$ is given by (\ref{frotate}) with $r_f=0$ and $l_f=\infty$. 
One can check that the $f$-metric is flat for this choice of parameters, although the line element is written in an unusual form. 
It can be obtained from the canonical Minkowski metric $ds^2_M = -dt^2 + dx^2 +dy^2 +dz^2$ by the coordinate change,
$t=v-r$, $x+iy = (r-ia)e^{i\phi}\sin\theta$, $z=r\cos\theta$, 
and by the subsequent replacement $r\to C r$, $v\to C v$, $a\to C a$.
In this case, the metric $g$ is given by (\ref{grotate}) where the condition (\ref{cndab}) and the first constraint of (\ref{cndK}) must be satisfied. 

\subsection{Numerical solutions \label{sec:num}}

So far, all the solutions we showed belong to the Kerr family, which are also solutions of GR. However, the absence of no-hair theorems in massive gravity and in particular the non-validity of Birkhoff's theorem for spherically symmetric spacetimes, suggests the existence of other solutions such as BHs endowed with massive spin-2 hair. Due to the complexity of the field equations~(\ref{Eg}) and~(\ref{Ef}), finding such solutions requires the use of numerical methods. A detailed study of those solutions was performed in Ref.~\cite{Volkov:2012wp} who also studied the case where both metrics are diagonal but not necessarily proportional. In particular, asymptotically AdS hairy BH solutions were shown to exist. This was further extended in Ref.~\cite{Brito:2013xaa} where a family of asymptotically flat hairy BHs was found. More recently the same techniques were used to find vacuum wormhole solutions~\cite{Sushkov:2015fma} in these theories.

All the spherically symmetric uncharged solutions we discussed in the previous section are related through a coordinate change with the Schwarzschild-(A)dS metric.
It is possible to find other solutions by considering static spherically symmetric solutions of the field equations~(\ref{Eg}) and~(\ref{Ef}), with the generic ansatz for the metrics given by
\beq
g_{\mu\nu}dx^{\mu}dx^{\nu}&=&-Q^2\, dt^2 + N^{-2}\, dr^2 + R^2 d\Omega^2\,,\label{ansatz_g}\\
f_{\mu\nu}dx^{\mu}dx^{\nu}&=&-a^2\, dt^2 + b^{2}\, dr^2 + U^2 d\Omega^2\,,\label{ansatz_f}
\eeq
where $Q\,,N\,,a\,,b$, $R$ and $U$ are radial functions. Gauge freedom allow us to reparametrize the radial coordinate $r$ such that $R(r)=r$. To simplify the field equations we also introduce the radial function $Y(r)$ defined as $b=U'/Y$, where $'\equiv d/dr$. After using the conservation condition~(\ref{bianchi1}) the problem can be reduced to a closed system of three coupled first-order ODE's for the functions $N$, $Y$ and $U$ (the derivation of these equations can be found in Ref.~\cite{Volkov:2012wp} and is also available online in a {\scshape Mathematica} notebook~\cite{webpage}):
\be
\left\{\begin{array}{l}
  N'=\mathcal{F}_1(r,N,Y,U,\mu,\kappa,\alpha_3,\alpha_4)\\
	Y'=\mathcal{F}_2(r,N,Y,U,\mu,\kappa,\alpha_3,\alpha_4)\\
	U'=\mathcal{F}_3(r,N,Y,U,\mu,\kappa,\alpha_3,\alpha_4).
    \end{array}\right.\label{eqs_Volkov}\,
\ee
The remaining two functions $Q$ and $a$ can then be evaluated from algebraic equations of the form:
\beq
Q^{-1}Q'&=&\mathcal{F}_4(r,N,Y,U,\mu,\kappa,\alpha_3,\alpha_4)\,,\label{eqs_Volkov_2a}\\
Q^{-1}a&=&\mathcal{F}_5(r,N,Y,U,\mu,\kappa,\alpha_3,\alpha_4)\,.\label{eqs_Volkov_2}
\eeq
Here we defined the parameter
\begin{equation}
	\mu^2=\frac{m^2}{2}\left(1+\frac{1}{\kappa}\right)\,,
\end{equation}
which, as we will see later, corresponds to the graviton mass in some backgrounds. 
As discussed previously, when $f_{\mu\nu}=C^2 g_{\mu\nu}$, we find the same solutions of GR, namely Schwarzschild-(A)dS, where the value of the cosmological constant will depend on $C$ through Eq.~(\ref{cnd2}). However, when we do not require the metrics to be proportional the complexity of the system of Eqs.~(\ref{eqs_Volkov}) requires them to be solved numerically. 

To solve these equations, appropriate boundary conditions at the event horizon $r_h$ must be imposed. For the spacetime to be smooth at the horizon, bidiagonal solutions must share the same horizon~\cite{Deffayet:2011rh}, which implies that $Q(r_h)=N(r_h)=Y(r_h)=a(r_h)=0$. Assuming a power-series expansion at the horizon of the form
\beq
N^2&=&\sum_{n\geq 1}a_n(r-r_h)^n, \quad Y^2=\sum_{n\geq 1}b_n(r-r_h)^n,\label{BCs1}\\
U&=&u\,r_h+\sum_{n\geq 1}c_n(r-r_h)^n\label{BCs3}\,.
\eeq
it follows that, for a given set of parameters of the theory, the solutions are parametrized by one single free parameter $u$. Fixing the value of $u=C$, with $C$ obtained from Eq.~(\ref{cnd2}) and integrating numerically the equations from the horizon $r=r_h$ up to infinity, the solution is given by Schwarzschild-(A)dS, with the cosmological constant fixed by Eq.~(\ref{cnd2})~\cite{Volkov:2012wp}. On the other hand more interesting non-trivial solutions appear when choosing $u=C+\delta u$.

\subsubsection{Asymptotically AdS hairy solutions.}

\begin{figure}[htb]
\begin{center}
\begin{tabular}{cc}
\epsfig{file=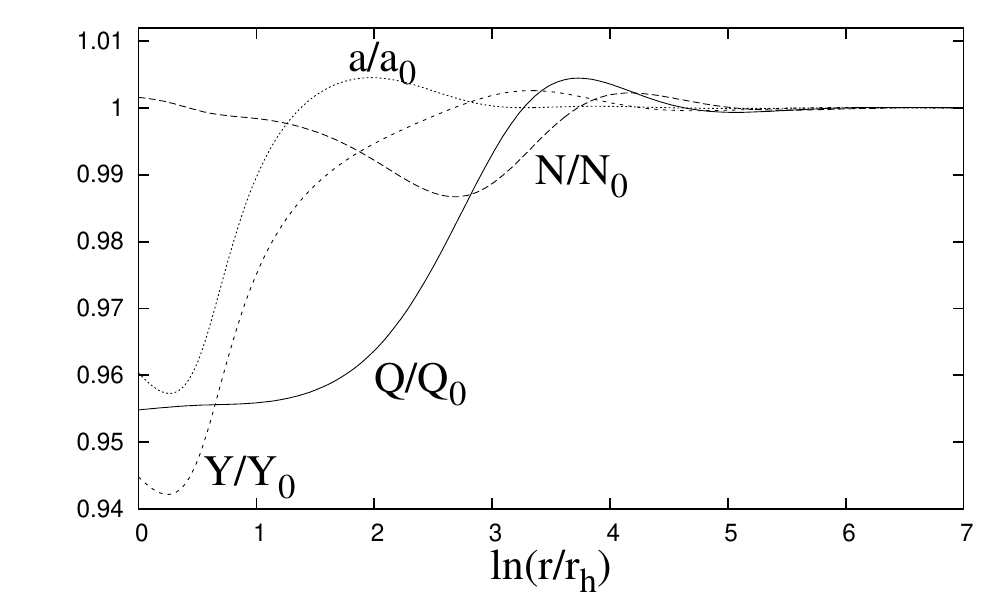,width=8cm,angle=0,clip=true}
\epsfig{file=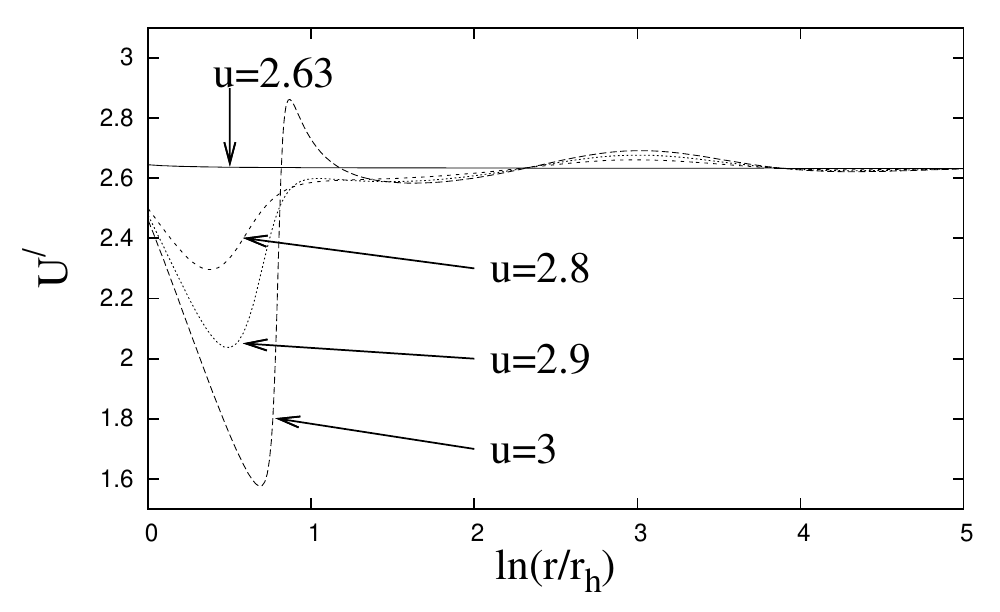,width=8.2cm,angle=0,clip=true}
\end{tabular}
\caption{Left: Example of metric functions for an hairy asymptotically AdS solution for $\alpha_3=-0.1$, $\alpha_4=-0.3$, and $\kappa=0.41$, where $N_0$, $Q_0$, $Y_0$ and $a_0$ correspond to the Schwarzschild-AdS solution with $u=2.6333$. Right: Function $U'(r)$ for different values of $u$. Taken from~\cite{Volkov:2012wp}.\label{ads_sols}}
\end{center}
\end{figure}

Using this method hairy deformations of the Schwarzschild-AdS geometry were found in Ref.~\cite{Volkov:2012wp}. These solutions approach AdS spacetime when $r\to\infty$, showing deviations from it close to the horizon. They exist for continuous small deformations $\delta u$ around the bi-Schwarzschild-AdS solution. An example of such solutions is shown in Fig.~\ref{ads_sols}.

Interestingly, when one takes the limit where the horizon radius vanish $r_h\to 0$, solutions with no horizon and purely made of massive field modes still exist~\cite{Volkov:2012wp}. These are globally regular solutions, including at $r=0$, and asymptotically approach AdS. More recently, in addition to these solutions, Ref.~\cite{Sushkov:2015fma} showed that for some discrete values of $u$ there also exist wormhole solutions which are asymptotically AdS. The existence of such solutions is related to the fact that the energy-momentum tensors~(\ref{Tmnexpl}) and~(\ref{Tmnexplf}) do not, in general, satisfy the null energy condition~\cite{Baccetti:2012re}. 

\subsubsection{Asymptotically flat hairy solutions.}

Unlike the AdS case, constructing asymptotically hairy flat BHs turns out to be much more complicated. In fact, as was pointed out in Ref.~\cite{Volkov:2012wp}, solutions which differ from Schwarzschild can only exist for discrete values of $u$ (fixing all the other parameters), unlike in AdS where solutions can be found varying $u$ continuously.

Those were found in Ref.~\cite{Brito:2013xaa}, with the solutions having an explicit ``Yukawa'' behavior when $r\to \infty$:
\beq
N&=&1-\frac{C_1}{2r}+\frac{C_2(1+r\mu)}{2r}e^{-r\mu}\,,\label{inf1}\\
Y&=&1-\frac{C_1}{2r}-\frac{C_2(1+r\mu)}{2r}e^{-r\mu}\,,\label{inf2}\\
U&=&r+\frac{C_2(1+r\mu+r^2\mu^2)}{\mu^2 r^2}e^{-r\mu}\,,\label{inf3}
\eeq
where $C_1$ and $C_2$ are integration constants. Note that the Yukawa behavior at infinity justifies identifying the parameter $\mu$ with the graviton mass. The value of $u$ for these solutions can be found fixing the values of $\mu$, $\alpha_3$, $\alpha_4$ and $\kappa$, integrating numerically from the horizon with the boundary conditions~(\ref{BCs1})--(\ref{BCs3}), and then shooting for the value of $u$ such that the solution matches the asymptotic behavior~(\ref{inf1})--(\ref{inf3}). {\scshape Mathematica} notebooks to generate these hairy solutions are available online~\cite{webpage}.

\begin{figure}[htb]
\begin{center}
\begin{tabular}{ccc}
\epsfig{file=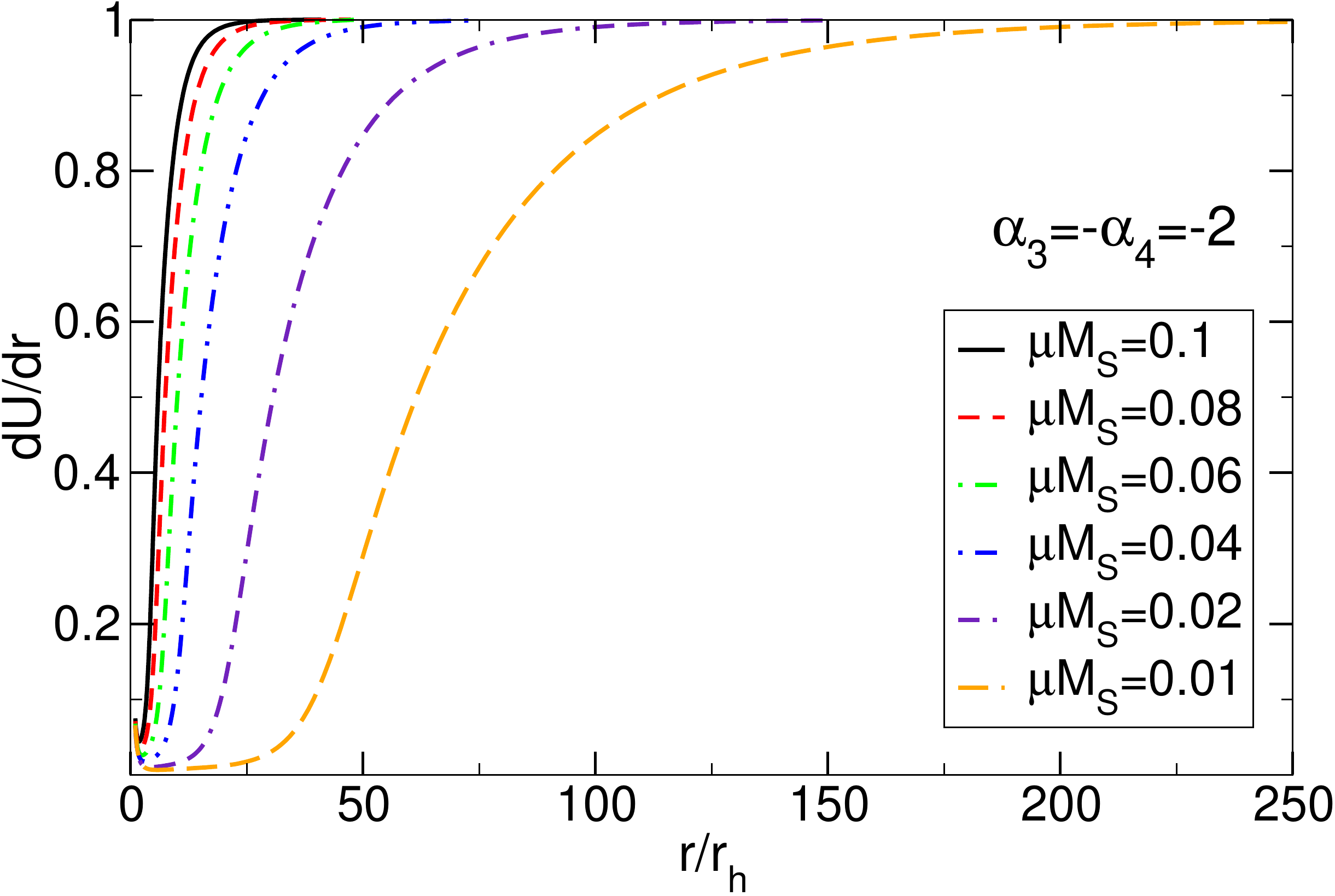,width=5cm,angle=0,clip=true}
\epsfig{file=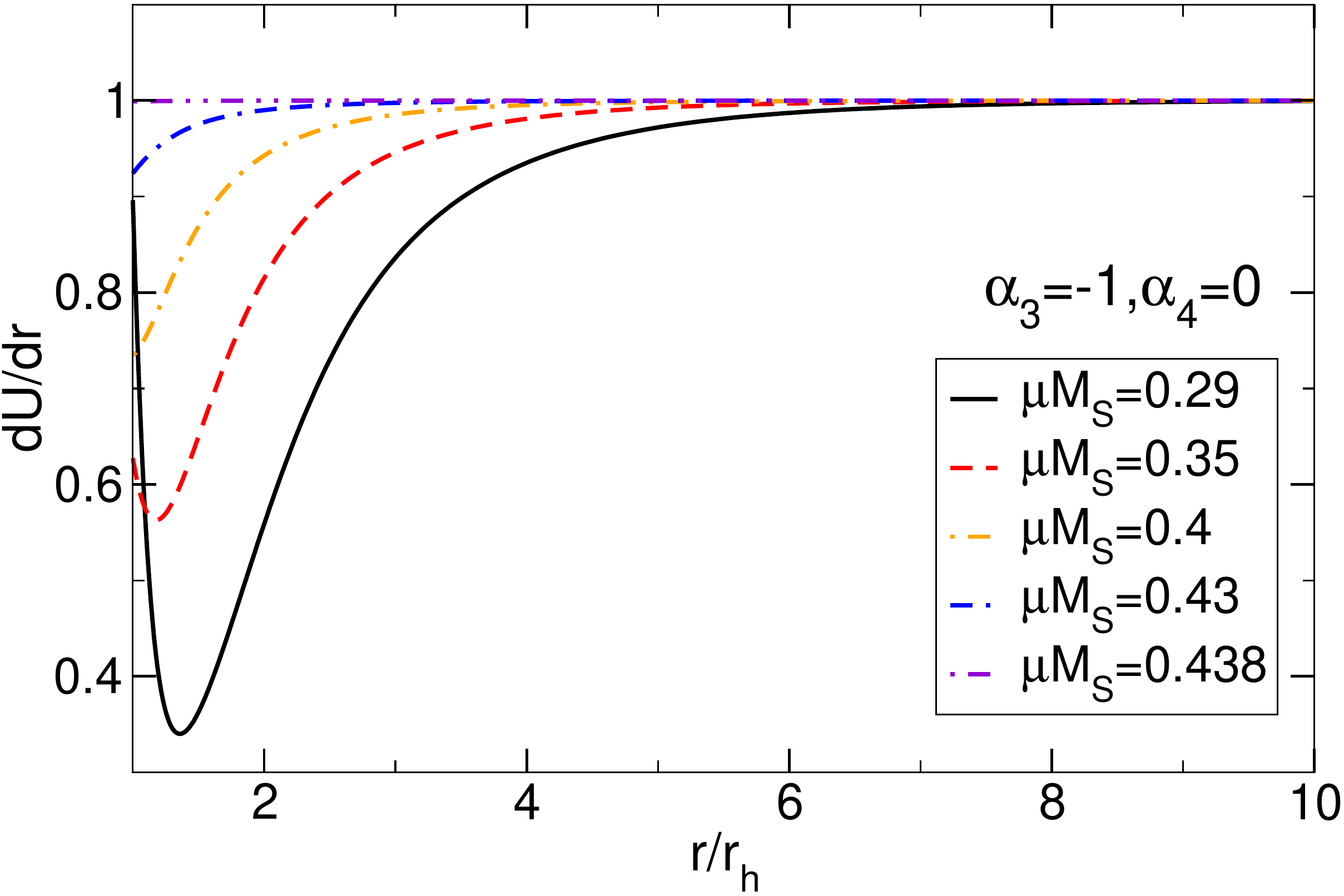,width=5cm,angle=0,clip=true}
\epsfig{file=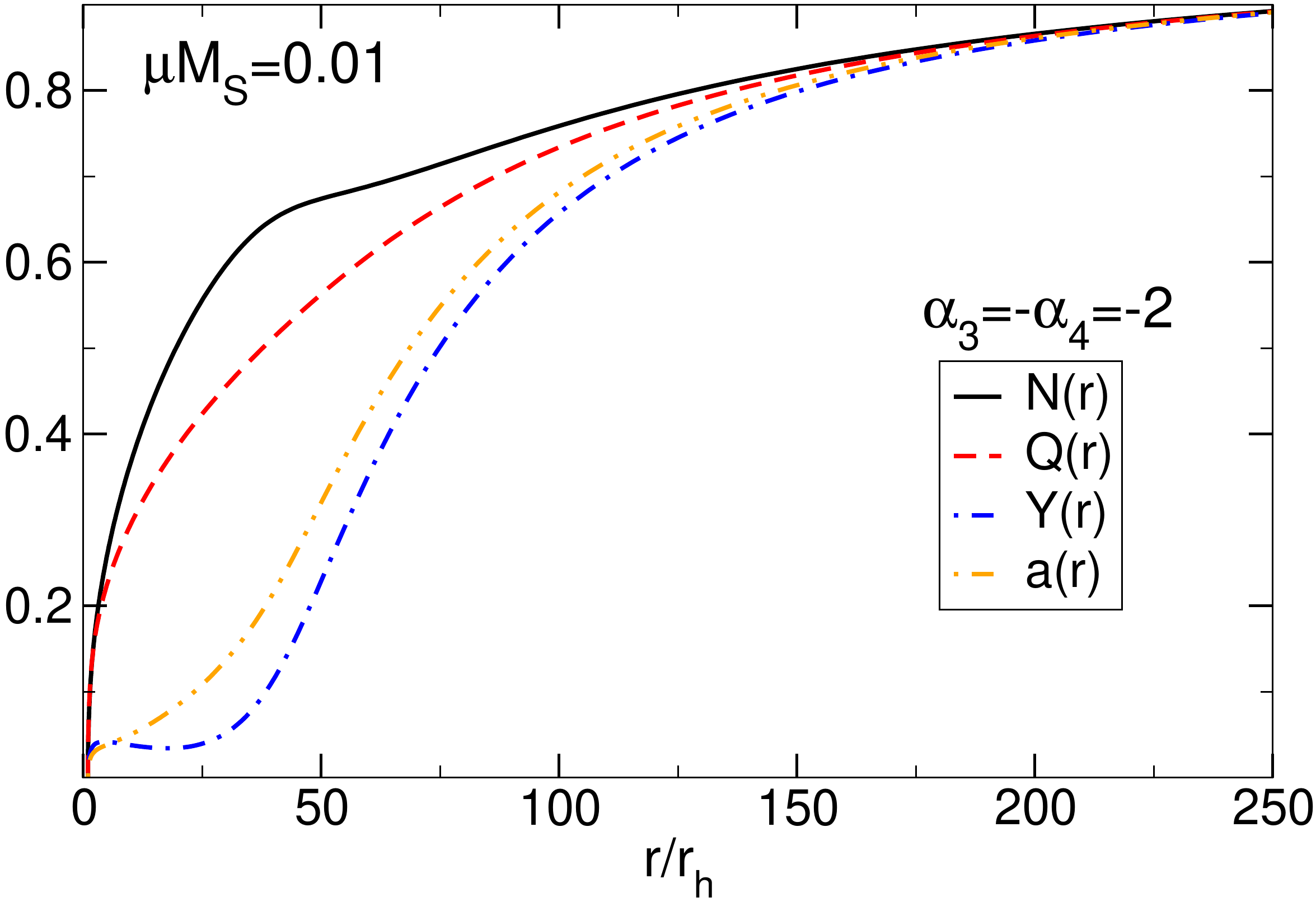,width=5cm,angle=0,clip=true}
\end{tabular}
\caption{Example of solutions for different values of the mass coupling  $\mu M_S$. The behavior is similar for any value $\alpha_3$ and $\alpha_4$ near the threshold $\mu M_S\sim 0.438$ but for small $\mu M_S$ it can be very different depending on the specific values of the parameters. Left panel: $\alpha_3=-2\,,\alpha_4=2$ and $\kappa=1$. Center panel: $\alpha_3=-1\,,\alpha_4=0$ and $\kappa=1$. Right panel: Metric functions for $\mu M_S=0.01$, $\alpha_3=-2\,,\alpha_4=2$ and $\kappa=1$. Taken from~\cite{Brito:2013xaa}.\label{fig:du}}
\end{center}
\end{figure}

A trivial solution for any value of $\mu$, $\alpha_3$ and $\alpha_4$ is obtained when $u=1$, and it corresponds to the two metrics being equal and described by the Schwarzschild solution. On the other hand, for $u\neq1$ there are also regular, asymptotically flat BHs endowed with a nontrivial massive graviton hair. The most important result of these studies is that hairy solutions exist near the threshold $\mu M_S\lesssim 0.438$ for {\it any} value of $\alpha_3,\alpha_4$ \footnote{The parameter $u=U(r_h)/r_h$ remains invariant under rescaling transformations. This can used to express all dimensionful quantities in terms of the mass of a Schwarzschild BH with horizon $r_h$, i.e. $M_S=r_h/2$.}. As we will discuss in Sec.~\ref{sec:radial} this is related to an instability of the bidiagonal Schwarschild solution found at the linear level for massive graviton masses satisfying precisely this bound~\cite{Babichev:2013una,Brito:2013wya}. The existence of this instability was in fact what prompted the authors of Ref.~\cite{Brito:2013xaa} to search for hairy solutions. At the threshold $\mu M_S\sim 0.438$ the hairy BH solution merges with the Schwarzschild solution, and above it the only bidiagonal solution seems to be the bi-Schwarschild one, which is consistent with the fact that this solution is linearly stable in this regime (see Sec~\ref{sec:radial} for more details). Examples of solutions for different choices of $\alpha_3$ and $\alpha_4$ are shown in Fig.~\ref{fig:du}.

\begin{figure}[htb]
\begin{center}
\begin{tabular}{c}
\epsfig{file=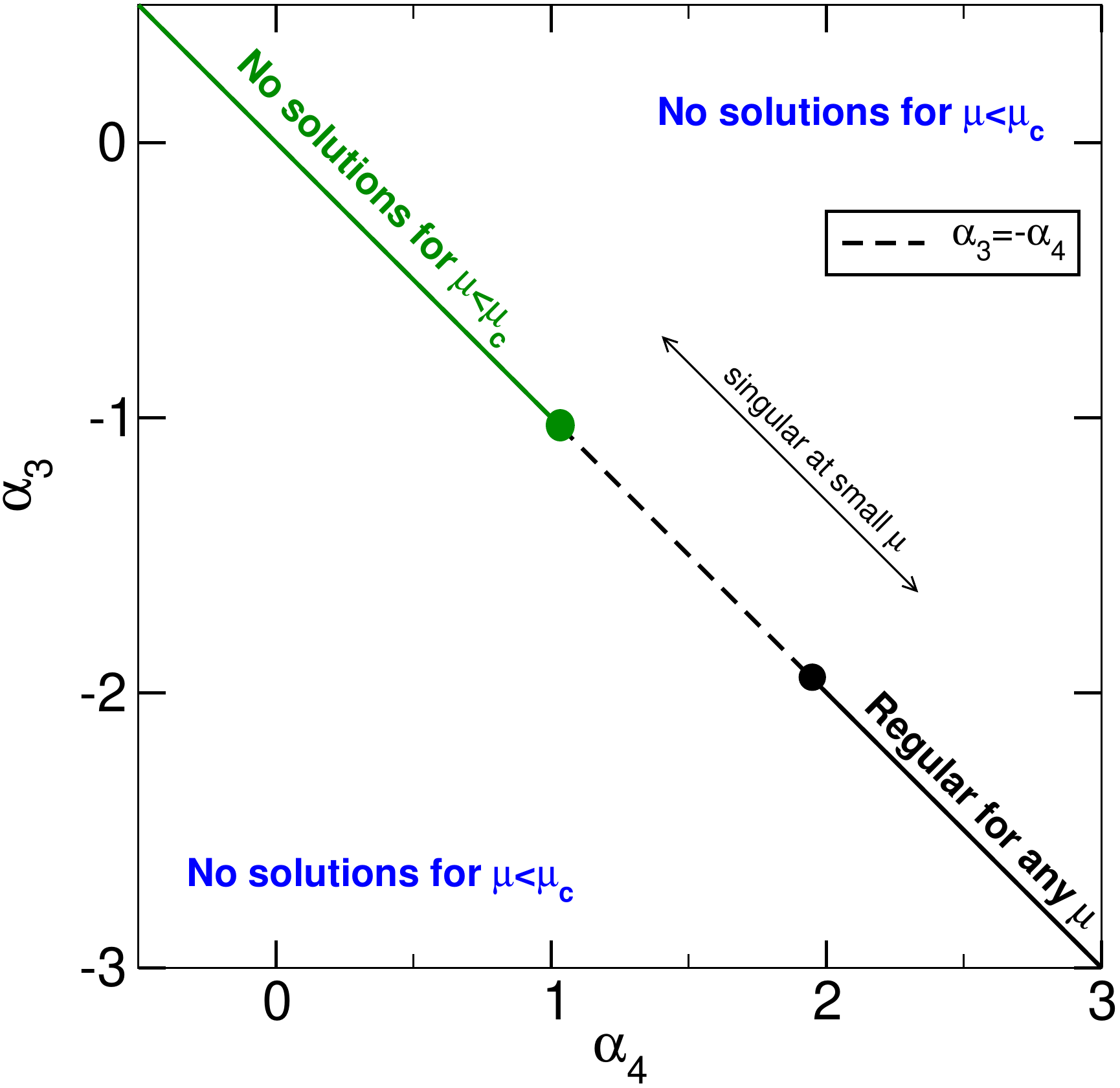,width=7.5cm,angle=0,clip=true}
\end{tabular}
\caption{Conjectured diagram of the parameter space for BHs with massive graviton hair in bimetric massive gravity. (i) $\alpha_3\neq-\alpha_4 \cup \alpha_3=-\alpha_4\lesssim -1$ -- The solutions stop to exist below a cutoff $\mu_c M_S$ (which depends on the parameters);
(ii) $1\lesssim\alpha_3=-\alpha_4\lesssim -2$ -- The solutions disappear only near $\mu M_S\sim 0.01$ and are singular at small $\mu M_S$, because some component of the metric $f_{\mu\nu}$ is vanishing where the metric $g_{\mu\nu}$ is regular; (iii) $\alpha_3=-\alpha_4\gtrsim -2$ -- The solutions exist for arbitrarily small $\mu M_S $ and are nonsingular. Taken from~\cite{Brito:2013xaa}.\label{fig:diagram}}
\end{center}
\end{figure}

The behavior at smaller $\mu M_S$ is more convoluted as it depends strongly on the nonlinear terms of the potential~(\ref{potential}). In fact, the solutions were found to stop to exist below a parameter dependent cutoff $\mu_c M_S$, and thus for some choices of the parameters $\alpha_3$ and $\alpha_4$, hairy BHs with arbitrarily small $\mu M_S$ seem to be excluded.
For graviton masses $\mu$ of the order of the Hubble scale this would mean that for these parameters only cosmologically large BHs exist, thus unlikely to be astrophysically relevant. 
Through a detailed numerical analysis of the two-dimensional parameter space $(\alpha_3\,,\alpha_4)$, these solutions were conjectured to follow the classification summarized in Fig.~\ref{fig:diagram}. However, it is unclear if for some parameters the solutions cease to exist when $\mu M_S\to 0$ or if they simply become extremely difficult to find numerically. In fact, in addition to the solution with asymptotic behavior~(\ref{inf1})--(\ref{inf3}), there is always another branch which diverges exponentially at spatial infinity. Thus, any small deviation from a regular solution leads to a singular behavior, making it numerically challenging to shoot for the correct solution when $\mu M_S\to 0$. It might be possible that using other numerical methods (such as a relaxation method) turns out to be more efficient. Therefore, it is unclear if the classification of Fig.~\ref{fig:diagram} is correct and further study is needed.

%

%

Finally, the above picture was found to break down in the limit where one of the metrics is taken to be a non-dynamical Schwarzschild metric ($\kappa \gg 1$). In this case the  numerical search suggest that, for any choice of $\alpha_3$ and $\alpha_4$, hairy BH solutions exist near the threshold $\mu M_S\lesssim 0.438$ but they do not exist for arbitrarily small $\mu M_S$.

To close this Section let us point out that asymptotically de-Sitter hairy BHs have, so far, not been found. The study of Ref.~\cite{Volkov:2012wp} seems to indicate that any small deviation from the Schwarzschild-de Sitter solution develops a curvature singularity at finite proper distance from the event horizon. However, whether a discrete set of solutions exist, such as the ones discussed in this Section, is still unknown. As in the asymptotically flat solutions discussed above, if any small deviation from a regular solution leads to a singular solution, those would be very difficult to find using a shooting method. In this case, other methods might be more appropriate.

\subsubsection{Other special black hole solutions.}

Other less interesting and pathological BH solutions were also found in Ref.~\cite{Volkov:2012wp}. For a matter of completeness let us give here a brief description of these solutions.

\paragraph{Asymptotically $U,a$ black holes.}
A class of bidiagonal solutions exist where the metric functions $U$ and $a$ of the metric $f$, given by Eq.~(\ref{ansatz_f}), asymptotically go to a constant value, while the metric $g$ asymptotically resembles AdS at leading order in $1/r$ (but not at all orders). These solutions are generically obtained if one continuously varies $u$ in the boundary condition~(\ref{BCs3}) around the Schwarzschild solution $u=1$, and are in fact asymptotically degenerate since $\det(f)\to 0$ when $r\to\infty$ and thus not physically relevant.

\paragraph{Tachyonic black holes.}
A class of special solutions with unusual asymptotic behavior was also found in the massive gravity limit where the non-dynamical metric is taken to be Schwarzschild. In this case, solutions with the asymptotic behavior~(\ref{inf1})--(\ref{inf3}) exist which have an imaginary graviton mass $\mu$. This suggest that around these solutions linearized perturbations are tachyonic, signaling an instability.

\section{Perturbations of black holes in massive (bi)-gravity}
\label{PERTURBATIONS}

Besides the existence of BH solutions in massive gravity, a fundamental issue one should raise concerns the stability properties of these solutions. Perturbation theory allow us to understand these properties and can sometimes even predict the existence of new solutions. Recently, the first steps of this programme have been done in Refs.~\cite{Babichev:2013una,Brito:2013wya,Brito:2013yxa,Kodama:2013rea,Babichev:2014oua}. They studied linear perturbations of some of the BH solutions discussed in the previous sections.

Surprisingly, they showed that the bidiagonal Schwarzschild BH solution is unstable. This instability was first discovered in Ref.~\cite{Babichev:2013una} who showed that the mass term for a massive spin-2 field in this background is equivalent to the Kaluza-Klein momentum of a four-dimensional Schwarzschild BH extended into a flat higher dimensional spacetime. These black strings, solutions of GR, have been shown to be unstable against long-wavelength perturbations, the so-called Gregory-Laflamme instability~\cite{Gregory:1993vy}. This led the authors of Ref.~\cite{Babichev:2013una} to conclude that perturbations of the bidiagonal Schwarzschild BH would generically give rise to
a spherically symmetric instability. This was fully confirmed and extended to the bidiagonal Schwarzschild-de Sitter solution in Ref.~\cite{Brito:2013wya}.

Perturbations of the bi-Kerr BH solution~(\ref{grotate}) and~(\ref{frotate}), for $f_{\mu\nu}=g_{\mu\nu}$, were also studied in Ref.~\cite{Brito:2013wya}, where it was found that due to the superradiant scattering of bosonic fields with spinning BHs another kind of instability could be triggered. Due to the dissipative nature of the BH horizon and to the existence of negative-energy states in the ergoregion of a spinning BH, low-frequency $\omega$ monochromatic bosonic waves scattered off rotating BHs are amplified whenever the following condition is met~\cite{Teukolsky:1974yv},
\be\label{super_cond}
\omega<m \Omega_H\,,
\ee
where $\Omega_H$ is the angular velocity of the BH horizon and $m$ is an integer characterizing the azimuthal dependence of the wave. The extra energy deposited in the wavepacket's amplitude is extracted from the BH, which spins down in the process. If a confining mechanism is able to trap superradiant modes near the BH, this can trigger an instability (see e.g. Refs.~\cite{Cardoso:2013krh,Brito:2015oca} and references therein). It turns out that a massive bosonic field can naturally confine low-frequency radiation due to the Yukawa-like suppression of the field at large distances, $\sim e^{-\mu r}/r$. This instability was explicitly shown to occur for scalar (spin-0)~\cite{Detweiler:1980uk,Dolan:2007mj} and vector (spin-1)~\cite{Pani:2012bp,Pani:2012vp,Witek:2012tr} fields and Ref.~\cite{Brito:2013wya} showed that massive tensor (spin-2) fields would also be unstable.

In this section we will consider small metric perturbations of the form $g_{\mu\nu}=g^{(0)}_{\mu\nu}+h_{\mu\nu}$, where $g^{(0)}_{\mu\nu}$ denotes the background metric at zeroth order in perturbation theory and $h_{\mu\nu}$ denotes a small perturbation of the background metric. To study spin-2 perturbations $h_{\mu\nu}$ of a generic spherically symmetric spacetime one can decompose the perturbations in tensor harmonics and write them in Fourier space as (in spherical coordinates)~\cite{Regge:1957td}
\begin{equation}
\label{decom}
h_{\mu\nu}(t,r,\theta,\phi)=\sum_{l,m}\int_{-\infty}^{+\infty}e^{-i\omega t}\left[h^{{\rm axial},lm}_{\mu\nu}(\omega,r,\theta,\phi)
+h^{{\rm polar},lm}_{\mu\nu}(\omega,r,\theta,\phi)\right]d\omega\,.
\end{equation}
where $h^{{\rm axial},lm}_{\mu\nu}$ and $h^{{\rm polar},lm}_{\mu\nu}$ are explicitly given by
\begin{eqnarray}\label{oddpart}
&&h^{{\rm axial},lm}_{\mu\nu}(\omega,r,\theta,\phi) =\nn\\
&& \left(
  \begin{array}{cccc}
  0 & 0 & h^{lm}_0(r)\csc\theta\partial_{\phi}Y_{lm}(\theta,\phi) & -h^{lm}_0(r)\sin\theta\partial_{\theta}Y_{lm}(\theta,\phi) \\
  * & 0 & h^{lm}_1(r)\csc\theta\partial_{\phi}Y_{lm}(\theta,\phi) & -h^{lm}_1(r)\sin\theta\partial_{\theta}Y_{lm}(\theta,\phi) \\
  *  & *  & -h^{lm}_2(r)\frac{X_{lm}(\theta,\phi)}{\sin\theta} & h^{lm}_2(r)\sin\theta W_{lm}(\theta,\phi)  \\
  * & * & * & h^{lm}_2(r)\sin\theta X_{lm}(\theta,\phi)
  \end{array}\right)\,,
\end{eqnarray}
\begin{eqnarray}\label{evenpart}
&&h^{{\rm polar},lm}_{\mu\nu}(\omega,r,\theta,\phi)=\nn\\
&&\left(
  \begin{array}{cccc}
H_0^{lm}(r)Y_{lm} & H_1^{lm}(r)Y_{lm} & \eta^{lm}_0(r)\partial_{\theta}Y_{lm}& \eta^{lm}_0(r)\partial_{\phi}Y_{lm}\\
  * & H_2^{lm}(r)Y_{lm} & \eta^{lm}_1(r)\partial_{\theta}Y_{lm}& \eta^{lm}_1(r)\partial_{\phi}Y_{lm}\\
  *  & *  & r^2h_{\theta\theta} & r^2  G^{lm}(r)X_{lm}  \\
  * & * & * & r^2\sin^2\theta h_{\phi\phi}
  \end{array}\right)\,,
\end{eqnarray}
%
where asterisks represent symmetric components, $h_{\theta\theta}\equiv K^{lm}(r)Y_{lm}+G^{lm}(r)W_{lm}$, $h_{\phi\phi}\equiv K^{lm}(r)Y_{lm}-G^{lm}(r)W_{lm}$, $Y_{lm}\equiv Y_{lm}(\theta,\phi)$ are the scalar spherical harmonics and

\be
X_{lm}(\theta,\phi)=2\partial_{\phi}\left[\partial_{\theta}Y_{lm}-\cot\theta Y_{lm}\right]\,,
\ee
\be
W_{lm}(\theta,\phi)=\partial^2_{\theta}Y_{lm}-\cot\theta\partial_{\theta}Y_{lm}-\csc^2\theta\partial^2_{\phi}Y_{lm}\,.
\ee
The perturbation variables are classified as ``polar'' or ``axial'' depending on how they transform under parity inversion ($\theta\to \pi-\theta$, $\phi\to \phi+\pi$). Polar perturbations are multiplied by $(-1)^l$ whereas axial perturbations pick up the opposite sign $(-1)^{l+1}$ (see e.g. Ref.~\cite{Berti:2009kk} for further terminology used in the literature).

In general the angular and radial parts of the perturbation $h_{\mu\nu}$ will fully decouple, such that the radial components will satisfy a set ODE's, which together with suitable boundary conditions at the BH horizon and at spatial infinity define an eigenvalue problem for the frequency $\omega$. These boundary conditions impose that the eigenfrequencies are generically complex, $\omega=\omega_R+i\omega_I$~\cite{Berti:2009kk}. Through Eq.~(\ref{decom}), an instability corresponds to an eigenfrequency with $\omega_I>0$ with an instability time scale $\tau\equiv1/\omega_I$, while the case $\omega_I<0$ corresponds to stable modes that decay exponentially in time.

\subsection{Radial perturbations\label{sec:radial}}

Let us first consider radial perturbations of the spherically symmetric solutions~(\ref{gEF}), (\ref{fEF}), and for 
the sake of simplicity we will mostly consider the case of asymptotically flat spacetimes, $l_g = l_f\to \infty$, 
\begin{eqnarray}
ds_g^2 & = & -\left(1-\frac{r_g}{r} \right) dv^2 +2dvdr+r^2 d\Omega^2,\label{gEF0}\\
ds_f^2 & = & C^2\left[- \left(1-\frac{r_f}{r}  \right) dv^2 +2dvdr+r^2 d\Omega^2\right]. \label{fEF0}
\end{eqnarray}
We will consider both the bidiagonal and non-bidiagonal solutions. 
The bidiagonal case is realized for $r_g = r_f$ and either $C=1$, $\Lambda_g = \Lambda_f =0$ (the case of identical metrics, $f_{\mu\nu}=g_{\mu\nu}$); 
or  $f_{\mu\nu} = C^2 g_{\mu\nu}$ and with the following relation between $C$ and the parameters of the action~(\ref{dGTBETAa}) (cf. Sec.~\ref{ana_sol}),
\begin{eqnarray}\label{cnd20}
	\Lambda_g  =  \lambda_g,\,\, C^4 \kappa \Lambda_f  =  \lambda_f.
\end{eqnarray}
As we showed in Sec.~\ref{ana_sol}, there is another class of solutions following from the ansatz~(\ref{gEF0}), (\ref{fEF0}), the non-bidiagonal class, $r_g\neq r_f$. 
For this class of solutions the scale factor $C$ satisfies the relation~(\ref{cndab}), and the parameters of the action must be related via the same relation 
as for the bidiagonal case for $C\neq 1$, Eq.~(\ref{cnd20}).

The metric perturbations $h_{\mu\nu}^{(g)}$ and $h_{\mu\nu}^{(f)}$, corresponding to the $g$ and $f$ metrics, satisfy the linearized field equations
\begin{equation}\label{perteqs}
	\delta G^{\mu}_{\phantom{\mu}\nu}  =  
	m^2 \delta T^{\mu}_{\phantom{\mu}\nu}, \ \ \ 
	\delta \mathcal{G}^{\mu}_{\phantom{\mu}\nu}  =  \frac{m^2}\kappa \delta\left(
		 \frac{\sqrt{-g}}{\sqrt{-f}} \mathcal{T}^{\mu}_{\phantom{\mu}\nu}  \right)\ .
\end{equation} 

For spherically symmetric modes only the terms proportional to $Y_{lm}$ in the ansatz~(\ref{decom}) remain. Writing the perturbations in ingoing Eddington-Finkelstein coordinates (and changing the notation of the radial functions to avoid confusion) we then have
\begin{equation}\label{swave}
h^{\mu\nu}_{(g)}=e^{-i\omega v}\left(
  \begin{array}{cccc}
   h^{vv}_{(g)}(r) & h^{vr}_{(g)}(r)  & 0 & 0 \\
   h^{vr}_{(g)}(r) & h^{rr}_{(g)}(r)  & 0 & 0 \\
    0 & 0 &  \frac{h^{\theta\theta}_{(g)}(r)}{r^2} & 0 \\
   0 & 0 & 0 & \frac{h^{\theta\theta}_{(g)}(r)}{r^2 \sin^2\theta} \\
  \end{array}\right)\ ,
  \end{equation}
  \begin{equation}\label{swavef}
h^{\mu\nu}_{(f)}=  \frac{e^{-i\omega v}}{C^2} \left(
  \begin{array}{cccc}
   h^{vv}_{(f)}(r) & h^{vr}_{(f)}(r)  & 0 & 0 \\
   h^{vr}_{(f)}(r) & h^{rr}_{(f)}(r)  & 0 & 0 \\
    0 & 0 &  \frac{h^{\theta\theta}_{(f)}(r)}{r^2} & 0 \\
   0 & 0 & 0 & \frac{h^{\theta\theta}_{(f)}(r)}{r^2 \sin^2\theta} \\
  \end{array}\right)\ ,
  \end{equation}
where an overall $1/C^2$ factor for $f$-metric is introduced for convenience. 
Note that the advanced time $v$ is regular at the future horizon. 
Therefore we require the metric perturbations $h^{\mu\nu}_{(g,f)}(r)$  to be regular at $r=r_g$ and $r=r_f$. 

\subsubsection{Non-bidiagonal solutions.}

The calculation of the mass term in Eq.~(\ref{perteqs}) yields the remarkably simple expression,
\begin{equation}\label{nbd}\fl
\delta T^{\mu}_{\phantom{\mu}\nu} =
\frac{\mathcal{A} \left(r_g-r_f\right) }{4 r}\, e^{-i\omega v} 
\left(
\begin{array}{cccc}
 0 & 0 & 0 & 0 \\
 h_{(-)}^{\theta \theta } & 0 & 0 & 0 \\
 0 & 0 & \frac{h_{(-)}^{vv}}{2} & 0 \\
 0 & 0 & 0 & \frac{h_{(-)}^{vv}}{2}
\end{array}
\right),\,\,\,  
\delta\left( \frac{\sqrt{-g}}{\sqrt{-f}} \mathcal{T}^{\mu}_{\phantom{\mu}\nu}\right)=-\delta T^\mu_{\phantom{\mu}\nu},
\end{equation}
where,
\begin{equation}\label{AA}
\mathcal{A} = \frac{C^2\left(\beta(C-1)^2-1\right)}{C-1},
\end{equation}
and  we defined,
$$h^{\mu\nu}_{(-)} \equiv h^{\mu\nu} _{(g)}- C^2h^{\mu\nu}_{(f)}.$$ 
Notice that for the above definition, e.g. $h^{vv}_{(-)}(r)=h^{vv}_{(g)}(r)-h^{vv}_{(f)}(r)$, taking into account the factor $1/C^2$ in the definition of $h^{\mu\nu}_{(f)}$.
In the bidiagonal case the calculations are much simpler (we consider this case in detail below), since the perturbed mass term has the Fierz-Pauli-like form,
\begin{eqnarray}
\delta T^\mu_{\phantom{\mu}\nu}  = \frac{C}{2}\left(\beta(C-1)^2-2\alpha(C-1)+1\right)  e^{-i\omega v} \left(\delta^{\mu}_\nu h^{(-)}-h^{\mu (-)}_{\phantom{\mu}\nu}\right),\label{bd1}\\
 \delta\left( \frac{\sqrt{-g}}{\sqrt{-f}} \mathcal{T}^{\mu}_{\phantom{\mu}\nu}\right)  =  -\delta T^\mu_{\phantom{\mu}\nu}.\label{bd2}
\end{eqnarray}
At the intersection of the two branches of solutions, 
for $r_g=r_f$ in (\ref{nbd}) and for $C$ fixed by  (\ref{cndab}) in (\ref{bd1}), 
we find $\delta T^{\mu}_{\phantom{\mu}\nu}=0$. This means that the perturbation equations are those of GR, as it is evident from Eqs.~(\ref{perteqs}).
In the case of different radii, $r_g\neq r_f$, and, in particular, for a flat metric $f$ the perturbations are different from GR, unless $\mathcal{A}=0$ in Eq.~(\ref{nbd}). 
The condition  $\mathcal{A}=0$, which reduces to $\beta=(C-1)^{-2}$ implies from (\ref{cndab}) $\beta=\alpha^2$. 
This particular choice of parameters, which was also studied in Ref.~\cite{Kodama:2013rea}, is not generic and corresponds to the enhanced symmetry BHs discussed in Sec.~\ref{ana_sol}.

The divergence of Eq.~(\ref{perteqs}) leads to the constraint,
\begin{equation}\label{conT}
 \nabla^\nu_{(f)}\delta \left(\frac{\sqrt{-g}}{\sqrt{-f}}\mathcal{T}^{\mu}_{\phantom{\mu}\nu}\right)\propto \nabla^\nu_{(g)}\delta T^{\mu}_{\phantom{\mu}\nu} =0.
\end{equation}
Applied to the non-bidiagonal case, this constraint gives,
\begin{equation}\label{divg}
\frac{\mathcal{A} \left(r_g-r_f\right) }{4 r^2}\, e^{-i\omega v}
\left\{-\left( r h_{(-)}^{\theta \theta}\right)', h_{(-)}^{vv},0,0\right\}  =0,
\end{equation}
which  immediately leads to the following simple condition on the components of the metric perturbations,
\begin{equation}\label{constnbd} 
h^{vv}_{(-)} =0 ,\; h^{\theta\theta}_{(-)} = \frac{c_0}r,
\end{equation} 
where $c_0$ is an integration constant. 
Using (\ref{constnbd}) back in (\ref{nbd}) one can see that there is only one nontrivial component of the matrix $\delta T^{\mu}_{\phantom{\mu}\nu}$, namely, 
$\delta T^{r}_{\phantom{r}v}\propto h_{(-)}^{\theta \theta }$.

In the non bidiagonal case there is no simple way to separate the gauge-invariant metric components from the GR part. 
However, thanks to the extremely simple form of (\ref{nbd}) and the constraints (\ref{constnbd}) it is possible to obtain an analytical solution of the perturbation equations 
(in contrast to the bidiagonal case, where the equation (\ref{GL}) must be solved numerically, as discussed below). 
Note that once (\ref{constnbd}) is substituted into (\ref{perteqs}), we get linear differential equations (equivalent to those of GR),
with the r.h.s. being source terms, given by the constraints~(\ref{constnbd}).
Therefore, general solutions for $h^{\mu\nu}_{(g)}$ and $h^{\mu\nu}_{(f)}$ 
contain a part that is gauge-dependent (same as in GR) plus a particular solution to the full equations, i.e.  
\begin{equation}
	h^{\mu\nu}_{(g,f)} =  h^{\mu\nu (g,f)}_{GR} + h^{\mu\nu (g,f)}_{(m)} . 
\end{equation}
The particular (gauge-invariant) solution  is  given by a single nonzero component for each metric perturbation
\begin{eqnarray}
 	h^{rr(g)}_{(m)} &=&  -\frac{\mathcal{A}(r_g-r_f) e^{-i\omega v}}{4i\omega } m^2  h_{(-)}^{\theta \theta }, \label{partsolg}\\
 	h^{rr(f)}_{(m)} &=& -\frac{h^{rr(g)}_{(m)}}{\kappa} . \label{partsolf}
 \end{eqnarray}
Both the homogeneous parts of the solution, $h^{\mu\nu (g)}_{GR}$ and $h^{\mu\nu (f)}_{GR}$, are, individually, pure gauge. 
They can be written as  
$ h^{\mu\nu}_{GR} =  -\nabla^\mu \xi^\nu -\nabla^\nu\xi^\mu$, where to recover~(\ref{swave}), we set $\xi^\mu  = e^{-i\omega v}\left\{ \xi^0(r),\xi^1(r),0,0 \right\}$. 
The relation between $\xi^\mu_{(g)}$ and $\xi^\mu_{(f)}$ is fixed by the constraint (\ref{constnbd}), namely, 
$\xi^0_{(f)} = \xi^0_{(g)} + c_1$, $\xi^1_{(f)} = \xi^1_{(g)} + {c_0}/2$, where $c_1$ is a second integration constant. 
Since we are free to choose the gauge, we can completely eliminate the homogeneous (GR) part of the $f$-metric perturbation, i.e. 
\begin{equation}\label{hf0}
h^{\mu\nu (f)}_{GR} =0, 
\end{equation}
leaving only the non-zero $h^{\mu\nu (g)}_{GR}$,
\begin{equation}\label{resexp}
h^{\mu\nu (g)}_{GR}=e^{-i\omega v}
\left(
\begin{array}{cccc}
0 & -i\omega  c_1 & 0 & 0 \\
 -i\omega  c_1 & -c_0 \left(i\omega + \frac{r_g}{2 r^2} \right) & 0 & 0 \\
 0 & 0 & c_0 r^{-3} & 0 \\
 0 & 0 & 0 & c_0 \csc ^2(\theta ) r^{-3}
\end{array}
\right).
\end{equation} 
Alternatively, one can choose the gauge such that $h^{\mu\nu (g)}_{GR} =0$, but $h^{\mu\nu (f)}_{GR} \neq 0$. It is in general, however, 
not possible to set both of the homogeneous parts to zero. 

Note that static perturbations $\omega=0$ seem to be excluded due to the term $\omega^{-1}$ in (\ref{partsolg}), (\ref{partsolf}),
unless we take $c_0\sim \omega$. However, in the latter case, $h^{\mu\nu (f)}_{GR}$ vanishes and 
the only nonvanishing contribution left in $h^{rr}$ is~$\sim 1/r$, which describes the same non-bidiagonal solutions with properly redefined $r_g$ and $r_f$.
We can therefore exclude the existence of another branch of static solutions close to this family.

Taking $\omega =i\Omega$, with real positive $\Omega$, one notices, that the metric perturbations are regular at the horizon, 
but they are not regular at infinity. This observation rules out unstable spherically symmetric modes for non-bidiagonal solutions.
Moreover, even for real $\omega$, with ``usual'' boundary conditions for this type of problem, 
namely, outgoing waves at infinity and ingoing waves at the horizon, the only solution is  $c_0=c_1=0$. 
On the other hand, non-trivial purely ingoing waves do exist (as well as purely outgoing), specified by two integration constants $c_0$ and $c_1$,
which indicates the presence of the helicity-0 mode.

Once the result in the bi-metric theory is obtained, 
it is not difficult to adopt it to the case of the original dRGT theory.  
For $f_{\mu\nu}=\eta_{\mu\nu}$ we have $h^{\mu\nu(f)} = 0$, while 
 Eqs.~(\ref{partsolg}) and (\ref{resexp}) give the solution for perturbations of $g_{\mu\nu}$. 
 As in the case of the bi-metric theory, this solution does not have the correct behavior at $r\to \infty$, and therefore must be excluded.

Let us also comment that the scale factor $C$ needs to be fine-tuned in the case of non-bidiagonal solutions by the condition~(\ref{cndab}).
This particular choice of $C$ implies the absence of the scalar mode for the bi-flat spacetime, 
i.e. the helicity-0 mode is strongly coupled.
Indeed, from Eqs.~(\ref{partsolg}), (\ref{partsolf}) and (\ref{constnbd}) one finds 
for  $r\to\infty$, that the ``massive'' part of the perturbations $h^{\mu\nu(g,f)}_{(m)}$ disappears.
It is interesting that the scalar degree of freedom seems to be restored for non-zero curvature.

\subsubsection{Bidiagonal solutions.}

In the bidiagonal case, the situation is very different. The constraint (\ref{conT}) requires $h^{\mu\nu}_{(-)}$ to be  traceless and divergence-free,
   \begin{equation}\label{constbd}
   \nabla_{\mu}h^{\mu\nu}_{(-)}=h_{(-)}=0.
   \end{equation}   
Taking into account Eqs.~(\ref{bd1}),~(\ref{bd2}) and the constraint~(\ref{constbd}), 
the perturbation equations (\ref{perteqs}) in the bidiagonal case can be rearranged to describe one equation 
for the mode $h^{(+)\mu}_{\phantom{(\pm)}\nu} \equiv h^{\mu(g)}_{\nu} + \kappa h^{\mu(f)}_{\nu} $, and another for the 
mode $h^{\mu\nu}_{(-)}$ defined earlier. 
The combination  $h^{(+)\mu}_{\phantom{(\pm)}\nu}$ follows the perturbed Einstein's equation of GR,
\begin{equation}
\delta G^{\mu}_{\phantom{\mu}\nu} + \kappa\, \delta \mathcal{G}^{\mu}_{\phantom{\mu}\nu} = 0,
\end{equation}
which gives no instability, since this mode is fully equivalent to a gravitational perturbation of a Schwarzschild BH in GR~(see e.g.~\cite{Berti:2009kk}).
On the other hand, the equation for $h^{\mu\nu}_{(-)}$ is the massive Lichnerowicz equation supplemented by two constraints, which after generalizing it to include the cosmological constant reads:
\be
\left\{
\begin{array}{l}
\label{GL}
 \Box h_{(-)}^{\mu\nu} + 2 R_{\sigma\mu\lambda\nu} h_{(-)}^{\lambda\sigma} = \mu^2 h_{(-)}^{\mu\nu}\, , \\
 \mu^2\nabla_{\mu}h_{(-)}^{\mu\nu}=0\,,\\
 \left(\mu^2-{2\Lambda}/{3}\right)h_{(-)}=0\,,
\end{array}\right.
\ee
where $\Lambda\equiv \Lambda_g=\Lambda_f$ and 
\begin{equation*}
	\mu^2=\frac{m^2}{2}\left(1+\frac{1}{\kappa}\right)C\left(\beta(C-1)^2-2\alpha(C-1)+1\right).
\end{equation*}
These equations can be shown to be consistent only if we assume the background to be a vacuum solution of Einstein's equations with a cosmological constant $\Lambda$, 
so that $R=4\Lambda$, $R_{\mu\nu}=\Lambda g_{\mu\nu}$~\cite{Buchbinder:1999ar}. It should be noted  that under the infinitesimal coordinate change $x^\mu \to x^\mu + \xi^\mu$,
the equations of motion for the massive spin-2 perturbation $h^{(-)}_{\mu\nu}$ are not invariant, 
while for the massless spin-2 perturbation $h^{(+)}_{\mu\nu}$ this is a gauge transformation, leaving  the equations of motion unchanged. 
This is the reason why $h^{(+)}_{\mu\nu}$ does not have a propagating radial mode (and hence no instability): it can be simply gauged away, as in GR.

\begin{figure*}[htb]
\begin{center}
\begin{tabular}{c}
\epsfig{file=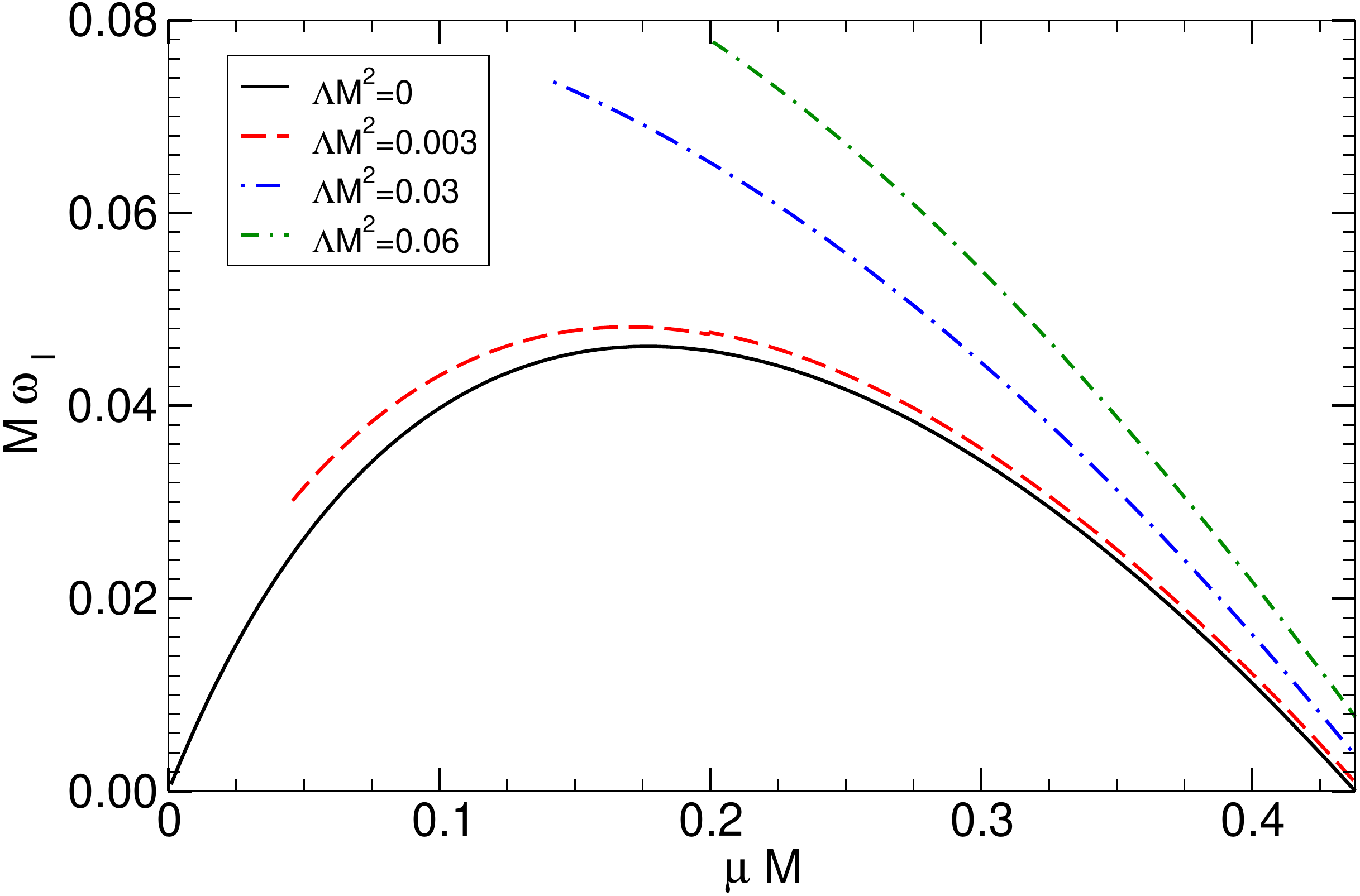,width=9cm,angle=0,clip=true}
\end{tabular}
\caption{Details of the instability of Schwarzschild (de Sitter) BHs against spherically symmetric polar modes of a massive spin-2 field. The plot shows the inverse of the instability timescale
$\omega_I=1/\tau$ as a function of the graviton mass $\mu$ for different values of the cosmological constant $\Lambda\equiv\Lambda_g=\Lambda_f$, including the asymptotically flat case $\Lambda=0$. Curves are truncated when the Higuchi bound is reached $\mu^2=2\Lambda/3$~\cite{Higuchi:1986py}. When this bound is saturated, the helicity-0 mode becomes pure gauge and the instability disappears~\cite{Brito:2013yxa}, while below this bound the mode becomes a ghost~\cite{Higuchi:1986py}. For any value of $\Lambda$, unstable modes exist in the range $0<M\mu\lesssim 0.47$, the upper bound being only mildly sensitive to $\Lambda$. Taken from~\cite{Brito:2013wya}.
\label{fig:GL}}
\end{center}
\end{figure*}

The system of Eqs.~(\ref{GL}) has been studied by Gregory and Laflamme~\cite{Gregory:1993vy}, although 
in a different context, not connected to massive gravity. They showed that this system admits the existence of unstable modes for 
\begin{equation}\label{inst}
0<\mu<\frac{O(1)}{r_g}\,.
\end{equation}
This led the authors of Ref.~\cite{Babichev:2013una} to the conclusion that the bi-Schwarzschild massive gravity solutions are unstable provided $\mu$ satisfies the condition~(\ref{inst}). 
In Ref.~\cite{Brito:2013wya} this instability was studied in detail also for the bi-Schwarzschild-de Sitter, and they showed that the unstable modes could be described by a single master equation given by
\begin{equation}
\label{evenl0}
\frac{d^2}{dr_*^2} \varphi_0 + \left[\omega^2-V_0(r)\right]\varphi_0=0\,,
\end{equation}
where $dr/dr_*=f\equiv 1-r_g/r$, $\varphi_0$ is a linear combination of polar functions in the decomposition~(\ref{decom}) and the potential is given by 
\begin{equation}
\eqalign{ 
 &V_0=\frac{1-2M/r-\Lambda /3\,r^2}{r^3 \left[2 M+r^3 \left(\mu ^2-2 {\Lambda/3}\right)\right]^2} \times\left\{8 M^3+12 M^2 r^3 \left(3 \mu ^2-8 {\Lambda/3}\right)\right.\nn\\
&\left.+r^7 \left(\mu ^2-2 {\Lambda/3}\right)^2 \left[6+r^2 \left(\mu ^2-2 {\Lambda/3}\right)\right]\right.\nn\\
&\left.-6 M r^4 \left(\mu ^2-2 {\Lambda/3}\right) \left[4+r^2 \left(3 \mu ^2-10 {\Lambda/3}\right)\right]\right\}\,,\label{V0dS} }
\end{equation}
where we defined the Schwarzschild radius $r_g=r_f=2M$, with $M$ the BH's mass (in units $G=c=1$).
This unstable mode is regular both at the horizon, where one must impose purely ingoing waves~\cite{Berti:2009kk}, and at infinity, where it behaves as
\be 
\varphi_0\sim e^{-\left(\sqrt{\mu^2-\omega^2}\right)r}\,, \quad {\rm as}\quad r\to \infty\,. 
\ee
The eigenvalues $\omega=\omega_R+ i \omega_I$ for these modes can be computed by numerically integrating Eq.~(\ref{evenl0}) with the appropriate boundary conditions. The result as a function of the mass coupling $M\mu$ is shown in Fig.~\ref{fig:GL}. The eigenfrequencies are characterized by a purely imaginary ($\omega=i\omega_I$), positive component ($\omega_I>0$), only exist for $M\mu\lesssim 0.43$ and for low masses (and $\Lambda=0$) behave as $\omega_I \sim 0.7\mu$.

Interestingly, for values of $M$ and $m$ that are phenomenologically relevant, namely considering the graviton mass to be of the order of the Hubble constant, $m\sim H\sim 10^{-33}{\rm eV}$, the mass coupling $\mu M$ is always well within the instability region (assuming $\kappa\sim 1$).  This implies that a graviton with such mass would trigger an instability for any Schwarzschild BH with mass smaller than $10^{22} M_\odot$! 
On the other hand, for the same typical value of the graviton mass, as it can be seen from Fig.~\ref{fig:GL}, 
the characteristic instability scale is of order of the inverse graviton mass, $\tau\sim m^{-1}$ (and is independent from the BH mass). 
Therefore for $m\sim H$, $\kappa\sim 1$ the characteristic instability time is of the order of Hubble time: 
for these scenarios, although the instability is present, it does not seem to be physically relevant for astrophysical black holes.

The scenario of instability depicted in the previous paragraph is valid as far as the dimensionless parameters, in particular $\kappa$, 
are of order of unity. However, it has been pointed out in \cite{Babichev:2013una} that the instability rate can be greatly enhanced for small $\kappa$.
Indeed, for small $\kappa$, but not extremely small, so that  (\ref{inst}) is satisfied, i.e.
$$(r_g m)^2 \lesssim \kappa \ll 1, $$
the instability rate is enhanced by a factor $\kappa^{-1/2}$.
For a given BH mass $M$, the maximum rate of instability is reached for $\kappa\sim (r_g m)^2$, for which the instability time scale is  $\tau \sim r_g$. 
This means that for $\kappa\sim (M_\odot m/M_P^2)^2$, 
the instability develops in $10^{-5}$ seconds for a solar mass black hole. 
On the other hand, in the limit $\kappa\to 0$, recently considered in \cite{Akrami:2015qga} for applications in cosmology,
the range of instability shrinks to zero.

As it is often the case, the existence of an instability hints at the existence of new equilibrium solutions branching off the instability threshold. In fact, the asymptotically flat hairy solutions discussed in Sec.~\ref{sec:num} correspond to this new family of solutions. However, the end-state of this instability is still unknown, and whether it drives the bidiagonal Schwarzschild solutions to these hairy solutions in some regions of the parameter space is unclear.

To close, let us note that there is also a region in the parameter space for which $\mu^2$ becomes negative. 
In this case, Eq.~(\ref{GL}) does not correspond to the case studied in~\cite{Gregory:1993vy}. 
However, the negative sign of $\mu^2$ signals an instability of a more dangerous type, 
namely the spin-0 part of the graviton becomes a ghost.
One can see this also from the Higuchi bound $\mu^2=2\Lambda/3$~\cite{Higuchi:1986py}, when the de-Sitter curvature goes to zero.

\subsection{Non-radial perturbations of proportional backgrounds}

Non-radial perturbations have, so far, only been studied for proportional BH solutions. These perturbations are governed by the field equations~(\ref{GL}) and were studied in detail in Refs.~\cite{Brito:2013wya,Brito:2013yxa}. Ref.~\cite{Brito:2013yxa} studied generic linear perturbations of the Schwarzschild-de Sitter solutions for the particular case where the Higuchi bound $\mu^2=2\Lambda/3$ is saturated and showed that the radial instability disappears, since the helicity-0 degree of freedom becomes pure gauge at the linear level (the so-called partially massless theory)~\footnote{However, there are strong indications to believe that at the full non-linear level this symmetry is always broken and the helicity-0 mode reappears~\cite{deRham:2013wv,Joung:2014aba,Deser:2013uy,Garcia-Saenz:2014cwa,Fasiello:2013woa}.}. On the other hand Ref.~\cite{Brito:2013wya} studied non-radial perturbations of asymptotically flat Schwarzschild and slow-rotating Kerr BHs, which we discuss below.

\subsubsection{Schwarzschild background.}

In a spherically symmetric background, when using the decomposition~(\ref{decom}), perturbations with opposite parity and different harmonic index $l$ decouple from each other. In addition, the radial and angular part of the field equations are separable such that the radial perturbation equations do not depend on the azimuthal number $m$. Inserting the ansatz~(\ref{decom}) into the system (\ref{GL}) two independent systems of ODEs can always be found, which can be schematically written as~\cite{Brito:2013wya}:
\begin{eqnarray}
 \mathbf{{\cal D}_A}\mathbf{\Psi_A}^l+\mathbf{V_A}\mathbf{\Psi_A}^l&=&0\,,\label{systA}\\
 \mathbf{{\cal D}_P}\mathbf{\Psi_P}^l+\mathbf{V_P}\mathbf{\Psi_P}^l&=&0\,,\label{systP}
\end{eqnarray}
where $\mathbf{{\cal D}_{A,P}}$ are second order differential operators, $\mathbf{V_{A,P}}$ are matrices, and $\mathbf{\Psi_{A}}$, $\mathbf{\Psi_{P}}$ are vectors of axial and polar functions, respectively. For $l \geq 2$, $\mathbf{\Psi_{A}}$ and $\mathbf{\Psi_{P}}$  are two and three-dimensional vectors, respectively. On the other hand for the dipole ($l=1$) mode, the angular functions $W_{lm}$ and $X_{lm}$ vanish and one is left with a single decoupled equation for the axial sector and a system of two equations for the polar sector. Finally, the monopole ($l=0$) mode does not exist in the axial sector since the angular part of the axial perturbations~(\ref{oddpart}) vanishes for $l=0$, while for the polar sector it can be reduced to Eq.~(\ref{evenl0}), which, as we discussed, allows for unstable modes. The full form of these equations can be found in Ref.~\cite{Brito:2013wya} and is available online in a {\scshape Mathematica} notebook~\cite{webpage}.
  
These equations admit the generic asymptotic solution at infinity $r \to \infty$,
\be
\Phi_j(r)\sim B_j e^{-k_{\infty} r}r^{-\frac{M(\mu^2-2\omega^2)}{k_{\infty}}}+C_j e^{k_{\infty} r}r^{\frac{M(\mu^2-2\omega^2)}{k_{\infty}}}\,,\quad r \to \infty\,,
\ee
where $k_{\infty}=\sqrt{\mu^2-\omega^2}$ and $\Phi_j$ schematically denotes the perturbation functions. This define two different families of physically motivated modes, which are distinguished according to how they behave at spatial infinity. The first family includes the standard quasinormal modes (QNMs), which corresponds to purely outgoing waves at infinity, i.e., they are defined by $B_j=0$~\cite{Berti:2009kk}. The second family includes quasibound states, defined by $C_j=0$. The latter correspond to modes spatially localized within the vicinity of the BH and that decay exponentially at spatial infinity (cf. Refs.~\cite{Dolan:2007mj,Rosa:2011my,Brito:2013wya,Pani:2012bp}). Imposing these boundary conditions jointly with purely ingoing waves at the horizon~\cite{Berti:2009kk},
\be
\label{BC_hor_Sch}
\Phi_j(r)\sim e^{-i\omega r_*}\,, \quad r_*\to -\infty\,,
\ee
where $dr/dr_*=f\equiv1-2M/r$, one can numerically solve for the eigenfrequency $\omega$ (see Ref.~\cite{Pani:2013pma} for a review on numerical techniques to solve this kind of problems). In Ref.~\cite{Brito:2013wya} these two different families of modes were studied in detail and no sign of instabilities were found, besides the spherically symmetric ($l=0$) mode discussed in the previous section.

\begin{figure}[htb]
\begin{center}
\epsfig{file=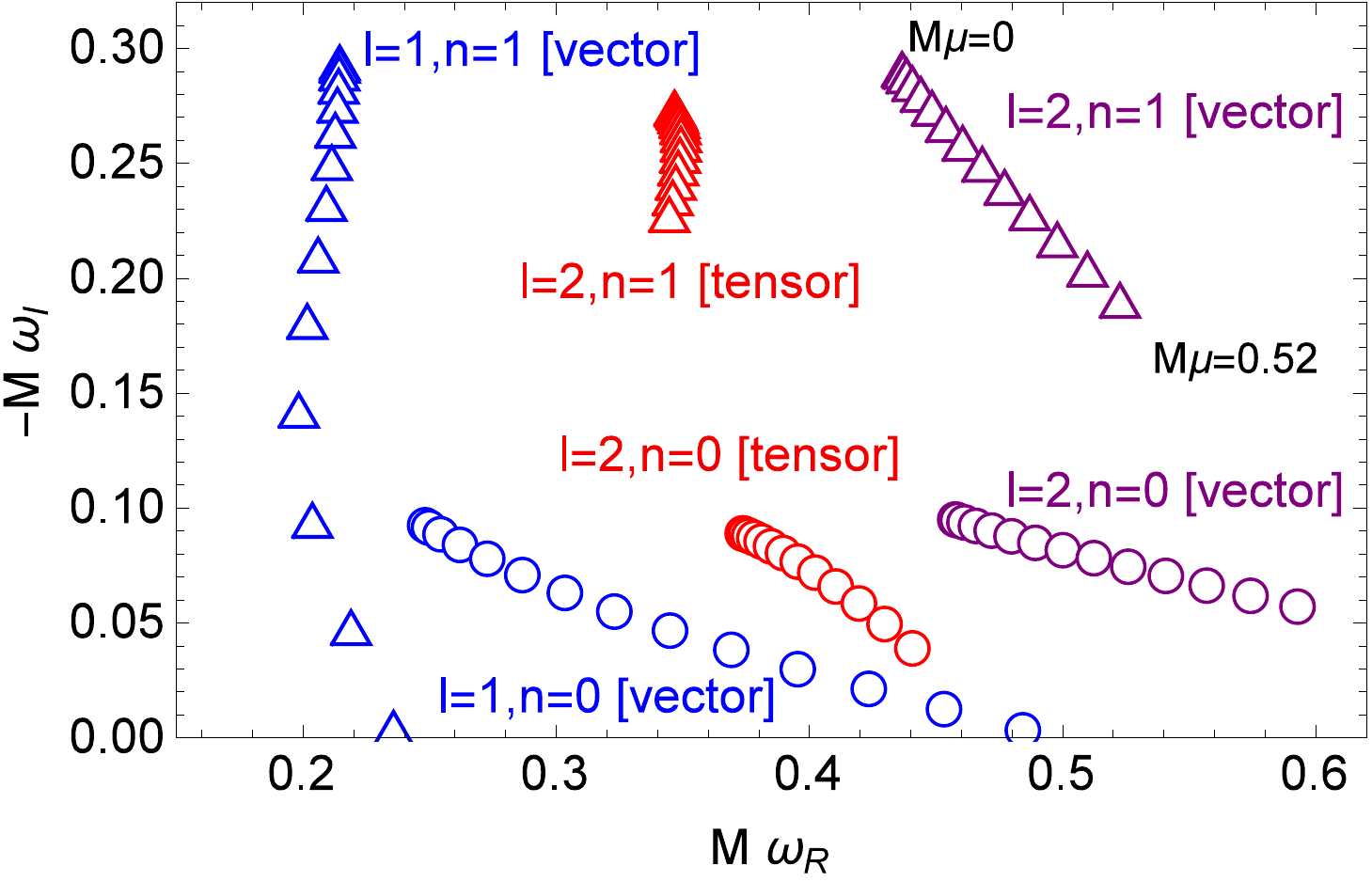,width=9cm,angle=0,clip=true}
\caption{QNM frequencies for axial $l=1,2$ modes, for a range of field masses $M\mu=0,0.04,\ldots,0.52$. Points with largest $|\omega_I|$ correspond to $\mu\to0$. The fundamental mode ($n=0$, circles) and the first overtones ($n=1$, triangles) are shown. In the massless limit the ``vector'' modes have the same QNM frequency as the electromagnetic perturbations of  a Schwarzschild BH in GR, and the ``tensor'' modes have the same QNM frequency as the massless gravity perturbations of a Schwarzschild BH. \label{fig:QNM} From Ref.~\cite{Brito:2013wya}.}
\end{center}
\end{figure}

Mainly due to the complexity of the polar equations, the QNMs were only studied for the axial sector. They found that the QNM spectrum is richer than in the GR case. Indeed, due to the additional degrees of freedom of the massive field, different families of modes exist which have no GR counterpart~\footnote{For the massive bi-gravity theory, as discussed in Sec.~\ref{sec:radial}, there are also modes coming from the field equations for $h^{(+)}_{\mu\nu} \equiv h^{(g)}_{\mu\nu} + \kappa h^{(f)}_{\mu\nu}$, which are fully equivalent to GR.}. These results are summarized in Fig.~\ref{fig:QNM}.

Quasi-bound states were studied in both sectors and were found to follow an hydrogenic spectrum in the small-mass limit ($M \mu\ll 1$):
\begin{eqnarray}
 \omega_R^2\sim&\sim&\mu^2\left[1-\left(\frac{M\mu}{l+n+S+1}\right)^2\right]\,,\label{wR}\\
 M\omega_I&\sim& -C_{Sl}(M\mu)^{4l+6+2S}\,,\label{wI}
\end{eqnarray}
where $n\geq0$ is the overtone number and $S$ is the polarization, while the coefficient $C_{Sl}$ depends on $S$ and $l$. The results above are valid for moderately large couplings $M\mu_V\lesssim 0.2$ and are in good agreement with what was previously found for massive spin-0 and spin-1 fields~\cite{Dolan:2007mj,Rosa:2011my,Pani:2012bp}. Note that these modes always have $\omega_I<0$ and thus they are stable and decay with a typical timescale $\tau=\omega_I^{-1}$. Interestingly, Eq.~(\ref{wR}) predicts a degeneracy for modes with the same value of
$l+n+S$ when $M\mu\ll 1$, which is akin to the degeneracy in the spectrum of the hydrogen atom. 

In addition to these modes, a new polar dipole ($l=1$) mode was found~\cite{Brito:2013wya}. This mode was shown to be isolated, does not follow the same small-mass behavior and does not have any overtone. For this mode, the real part is much smaller than the mass of the spin-2 field, and in the limit $M\mu\ll 1$ is very well fitted by
\beq
\omega_R/\mu&\approx& 0.72(1-M\mu)\,,\\\label{polar_di_Re}
\omega_I/\mu&\approx& -(M\mu)^{3}\,.\label{polar_di_Im}
\eeq
That this mode is different is not completely unexpected since in the massless limit it becomes unphysical. This peculiar behavior seems to be the 
result of a nontrivial coupling between the states with spin projection $S=-1$ and $S=0$. Besides that, this mode has the largest binding energy ($\omega_R/\mu-1$) among all couplings $M\mu$ for massive fields around Schwarzschild BHs, much higher than the ground states of the scalar, Dirac and vector fields.

\subsubsection{Slowly rotating Kerr BHs.}\label{sec:kerr}

Massless linearized gravitational fluctuations around the Kerr geometry were shown to be separable in GR by Teukolsky~\cite{Teukolsky:1973ha}. However, the Teukolsky formalism does not seem to be applicable for massive spin-2 perturbations governed by the field equations~(\ref{GL}). To handle the problem, Ref.~\cite{Brito:2013wya} considered an expansion in the rotation parameter $\tilde{a}\equiv a/M=J/M^2\ll 1$ where $J$ is the BH's angular momentum. Using the decomposition~(\ref{decom}) for the metric perturbations and
since the spacetime is not spherically symmetric, but instead axially symmetric, this leads to couplings between different $l$-modes and parity sectors, while perturbations with different values of $m$ fully decouple.

By expanding the Kerr background to first order in the spin\footnote[2]{As discussed in detail in~\cite{Pani:2013pma,Pani:2012bp}, a second-order calculation is needed to describe the superradiant regime in a self-consistent way, although a first-order computation turns out to be surprisingly accurate in all cases explored so far.}, the perturbation equations read schematically as (cf. Ref.~\cite{Pani:2012bp,Pani:2013pma,Brito:2013wya} for details)
\begin{eqnarray}
{\cal A}_{l}+\tilde a m \bar{\cal A}_{{l}}+\tilde a ({\cal Q}_{{l}}\tilde{\cal P}_{l-1}+{\cal Q}_{l+1}\tilde{\cal P}_{l+1})+{\cal O}(\tilde{a}^2)&=&0\,,\label{eq_axial}\\
{\cal P}_{l}+\tilde a m \bar{\cal P}_{{l}}+\tilde a ({\cal Q}_{{l}}\tilde{\cal A}_{l-1}+{\cal Q}_{l+1}\tilde{\cal A}_{l+1})+{\cal O}(\tilde{a}^2)&=&0\,,\label{eq_polar}
\end{eqnarray}
where, ${\cal Q}_l=\sqrt{\frac{l^2-m^2}{4l^2-1}}$ and the
coefficients ${\cal A}_l$ and ${\cal P}_l$ (with various
superscripts) are \emph{linear} combinations of axial and polar
perturbation variables, respectively. This system can always be written as a set of coupled second-order ODE's and when $\tilde{a}=0$, ${\cal A}_l$ and ${\cal P}_l$ reduces to Eqs.~(\ref{systA}) and (\ref{systP}), respectively. Furthermore, it can be shown that at first order in the spin, couplings between different values of $l$ do not affect the eigenspectrum and thus for numerical purposes one can fully separate the polar and axial sectors. 

As for the Schwarzschild case, at the horizon we must impose regular boundary conditions, which correspond to purely ingoing waves,
\be
\label{BC_hor_Kerr}
\Phi_j(r)\sim e^{-ik_H r_*}\,,
\ee
as $r_*\to -\infty$, where $k_H=\omega-m\Omega_H$, $\Omega_H=a/(2Mr_+)$ is the horizon angular velocity and $r_+=M+\sqrt{M^2-a^2}$ is the outer horizon of the Kerr geometry. 
All the polar and axial equations can be brought to a form such that the near-horizon solution is given by Eq.~(\ref{BC_hor_Kerr}). If $k_H<0$ an observer at infinity will see waves emerging from the BH~\cite{Teukolsky:1973ha}. This corresponds exactly to the superradiance condition, Eq.~(\ref{super_cond}). When modes satisfying this condition are confined in the vicinity of the BH, i.e. quasibound states, this leads to superradiant instabilities of bosonic massive fields. 

\begin{figure*}[htb]
\begin{center}
\epsfig{file=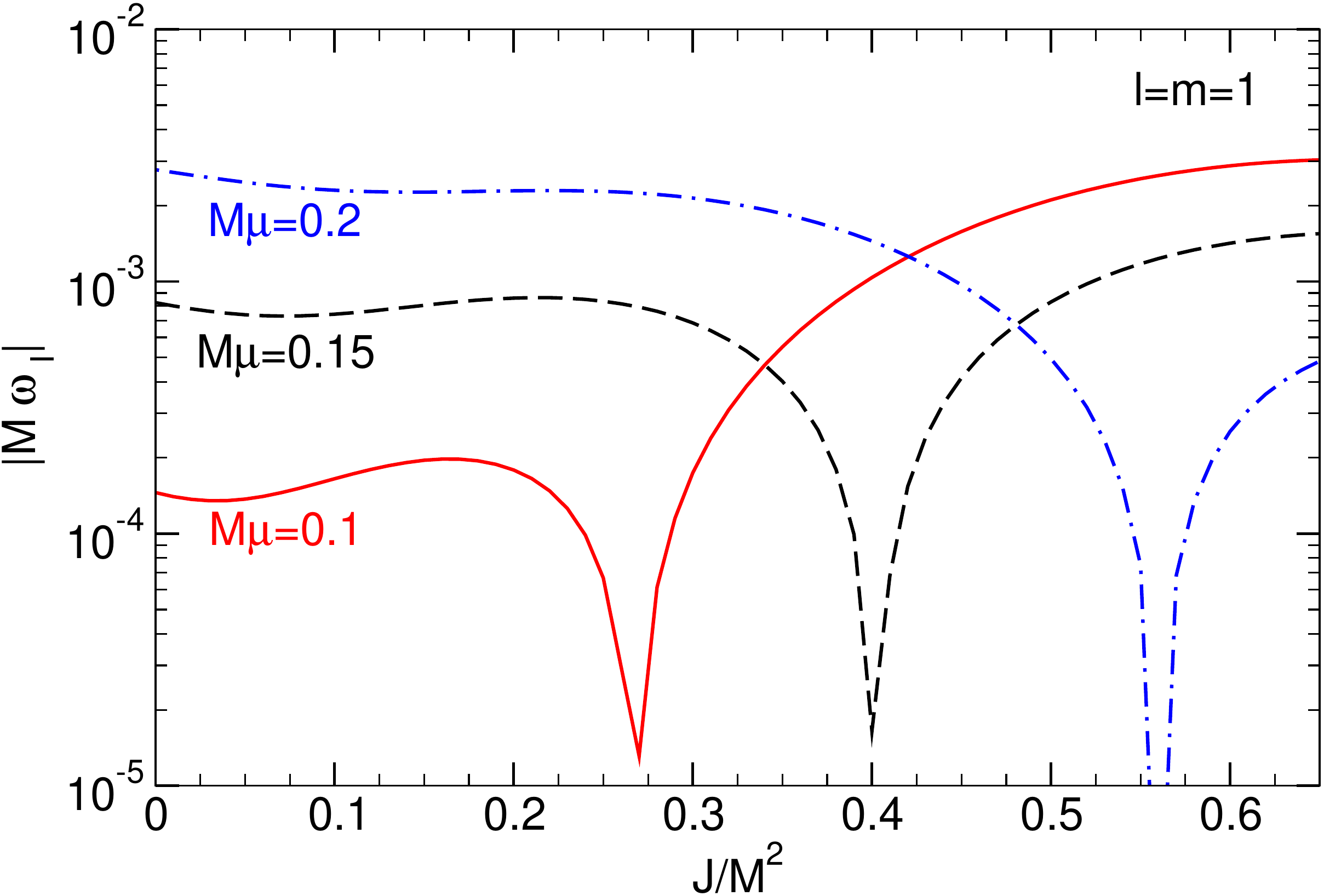,width=0.48\textwidth,angle=0,clip=true}
\epsfig{file=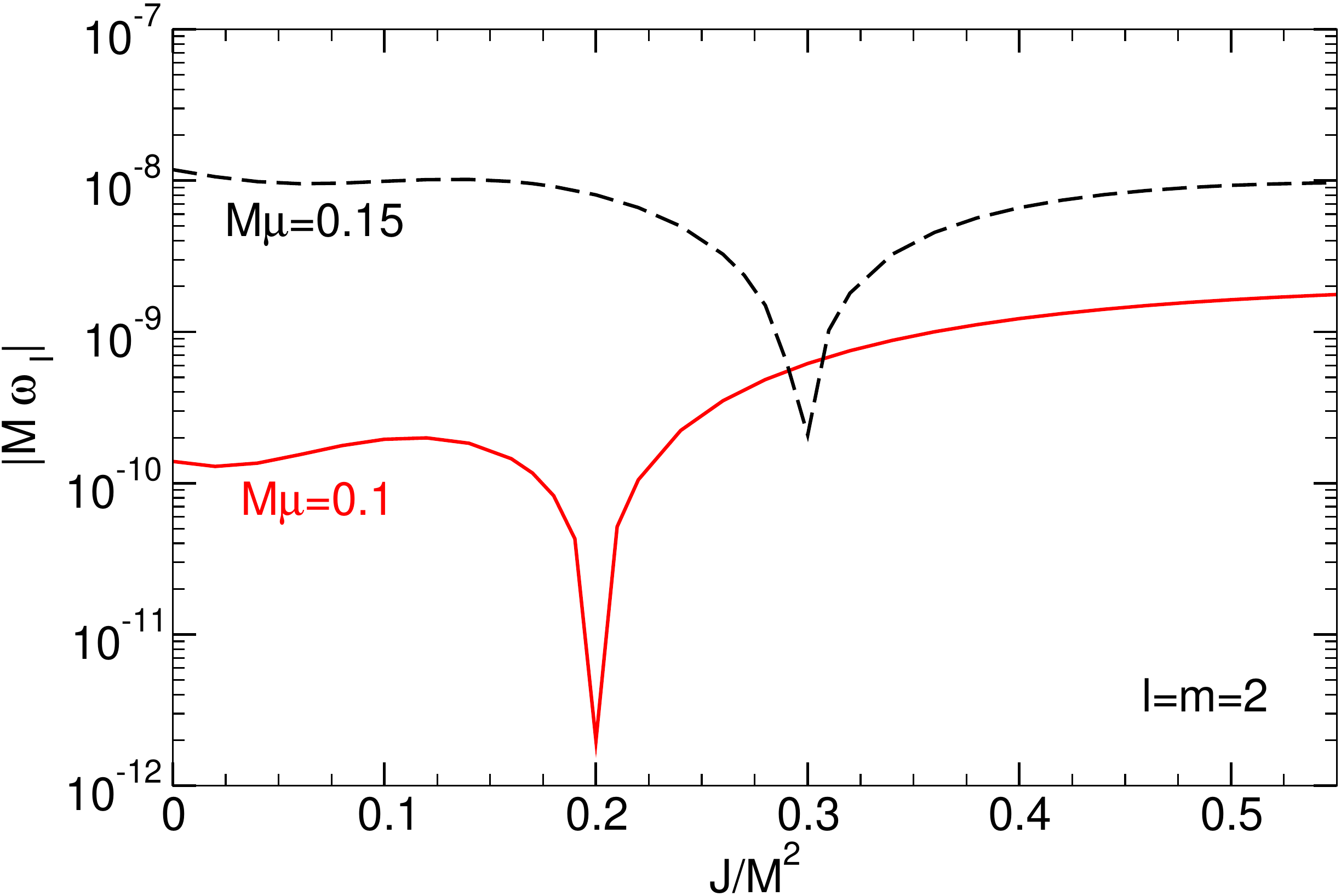,width=0.48\textwidth,angle=0,clip=true}
\caption{Absolute value of the imaginary part of the polar quasibound modes as a function of the BH rotation rate $J/M^2$ to first order in the spin for different values of $l$ and $m$ and different values of the mass coupling $\mu M$. 
Left panel: polar dipole mode for $l=m=1$. Right panel: polar mode $l=m=2$, $S=-2$. 
For any mode with $m\geq 0$, the imaginary part crosses the axis and become unstable when the superradiance condition $\omega_R<m\Omega_H$ is met. Taken from~\cite{Brito:2013wya}.
\label{fig:spin2}}
\end{center}
\end{figure*}

In the small $M\mu$ limit the eigenfrequencies of the quasibound states in a Kerr background can be obtained from Eqs.~(\ref{wR})--(\ref{polar_di_Im}) adding an extra factor $\left(2 r_+\mu-m a/M\right)$~\cite{Brito:2013wya}. Thus, in the superradiant regime, the imaginary part of the eigenfrequency $\omega_I$ changes sign, signaling an instability. An example of these modes is presented in Fig.~\ref{fig:spin2}, where it is shown that the decay rate of the dipole ($l=1$) polar mode is very large even for small couplings $M\mu$.
Indeed, for $M\mu\ll 1$ the time scale for this unstable mode is~\cite{Brito:2013wya}
\begin{equation}\label{polar_time}
 \tau\sim \frac{M(M\mu)^{-3}}{C_{{\rm polar}}(a/M-2r_+\omega_R)}\,,
\end{equation}
where $C_{{\rm polar}}\sim{\cal O}(1)$ and $\omega_R$ is given by Eq.~(\ref{polar_di_Re}). This is the shortest superradiant instability time scale of a Kerr BH known to date.

\subsection{Superradiant instability and bounds on the graviton mass}

The superradiant instabilities discussed above have been shown to have important astrophysical implications, which have been recently investigated in the contexts of testing stringy axions and ultralight scalars~\cite{Arvanitaki:2010sy,Kodama:2011zc,Arvanitaki:2014wva,Brito:2014wla}, to derive bounds on light vector fields~\cite{Pani:2012vp} and on massive gravitons~\cite{Brito:2013wya}.

\begin{figure}
\begin{center}
\epsfig{file=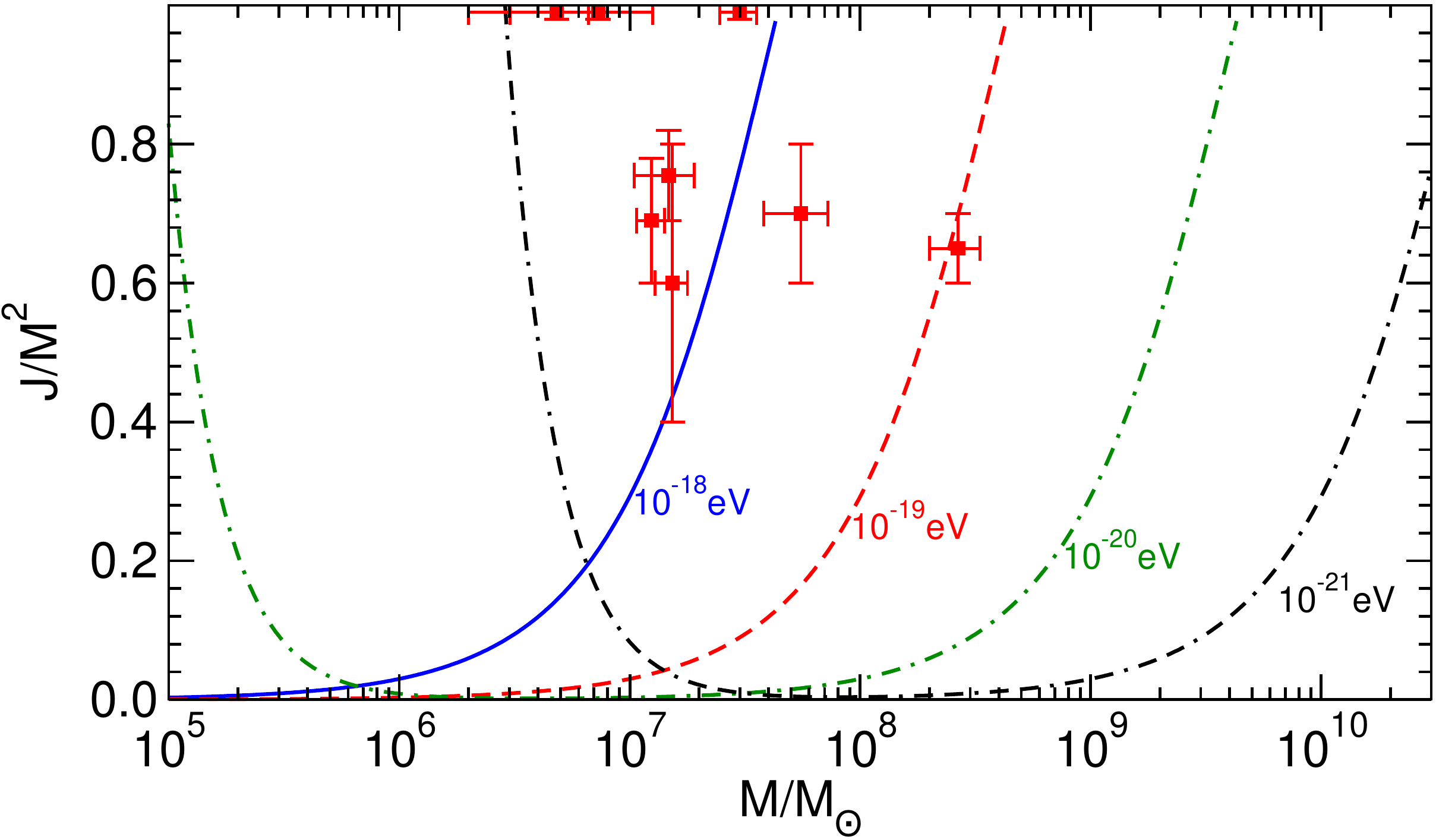,width=0.7\textwidth,angle=0,clip=true}
\caption{Contour plots in the BH Regge plane~\cite{Arvanitaki:2010sy}
  corresponding to an instability time scale given by Eq.~(\ref{polar_time}) shorter than a typical accretion time scale $\tau_{\rm Salpeter}\sim 4.5\times 10^7$ years, for different values of the massive spin-2 field mass
  $m_g\equiv{{\mu}}\hbar$ (see main text for details). The experimental points (with error bars) refer to the supermassive BHs
  listed in~\cite{Brenneman:2011wz}. Supermassive BHs lying above each
  of these curves would be unstable on an observable time scale, and
  therefore each point rules out a range of the graviton
  mass. Adapted from Ref.~\cite{Brito:2015oca}.\label{fig:bound}
\label{fig:ReggePlane}}
\end{center}
\end{figure}

Unlike the spherically symmetric instability (cf. Sec.~\ref{sec:radial}) for which the instability end-state is unclear, for the superradiant instability it is clear that the BH should spin-down until the superradiance condition~(\ref{super_cond}) is saturated~\cite{Brito:2014wla}. Thus, a solid prediction of BH superradiant instabilities is the existence of holes in the Regge plane~\cite{Arvanitaki:2010sy,Arvanitaki:2010sy,Brito:2014wla} (cf. Fig.~\ref{fig:ReggePlane}). Together with reliable spin measurements for massive BHs, this
can be used to impose stringent constraints on the allowed mass range
of ultralight bosons. 
These bounds follow from the requirement
that astrophysical spinning BHs should be stable, in the sense that
the superradiant instability time scale $\tau$ should be larger than some
observational threshold. For isolated BHs the most natural
observational threshold is the age of the Universe, $\tau_{\rm
  Hubble}=1.38\times 10^{10}$~years. However, for supermassive BHs, possible spin growth due to mergers with other BHs and/or
accretion also play an important role. The most likely mechanism to produce fastly-spinning BHs is
prolonged accretion~\cite{Berti:2008af}. Therefore, a conservative
assumption to estimate the astrophysical consequences of the
instability is to compare the superradiance time scale to the minimum
time scale over which accretion could spin up the BH. For simplicity we assume that
mass growth occurs via accretion at the Eddington limit, i.e. assuming equilibrium between the radiation pressure exerted on the matter surrounding the BH and the gravitational pull of the BH, so that the
BH mass grows exponentially with $e$-folding time given by the
Salpeter time scale $\tau_{\rm Salpeter}\sim 4.5\times 10^7$ years~\cite{Brito:2014wla}. Note that, assuming spherical symmetry, accretion at the Eddington rate sets an upper limit on the mass accretion rate.

We can quantify this statement by plotting exclusion regions in the BH Regge plane as shown in Fig.~\ref{fig:ReggePlane}. The contours correspond to an instability
time scale of the order of the Salpeter time for four different masses
of the bosonic field, considering the unstable mode with the largest growth rate, i.e., the polar dipole with an instability time scale given by Eq.~(\ref{polar_time}).
The plot shows that observations of supermassive BHs with $10^5M_\odot\lesssim
M\lesssim 10^{10}M_\odot$ spinning above a certain threshold would exclude a wide range of the massive spin-2 field mass. Note that the exclusion windows extend almost down to $J\sim 0$, and this feature is important given that current spin measurements might be affected by large systematics.

Nonetheless, it is clear from Fig.~\ref{fig:ReggePlane} that almost any supermassive BH spin measurement would exclude a
considerable range of masses. Since the only parameter that regulates the instability is the combination $\mu M$, similar exclusion plots exist in the region $M_\odot\lesssim M\lesssim 10^5 M_\odot$ for larger values of $\mu$. Thus, the best bound comes from the most massive BHs for which spin measurements are reliable, e.g. the BH candidate Fairall~9~\cite{Schmoll:2009gq}.

Using these arguments we can obtain the following bound~\cite{Brito:2013wya}\footnote{These bounds were obtained using a linearized analysis. By including the effects of gravitational-wave emission and gas accretion, Ref.~\cite{Brito:2014wla} showed that the linearized prediction should be corrected for massive scalar fields minimally coupled to GR, although such corrections would not affect the order of magnitude of these constraints. We expect that similar results should hold in theories of massive gravity.}:
\begin{equation}
m_g \lesssim 5\times 10^{-23} {\rm eV} \,, \label{bound_spin2}
\end{equation}
where $m_g\equiv \hbar \mu$. Note that, for a single BH observation, superradiant instabilities can only exclude a \emph{window} in the mass range of the field, as shown in Fig.~\ref{fig:ReggePlane}. Nonetheless, by combining different BH observations in a wide range of BH masses and joining the superradiant bounds with bounds coming from other experiments~\cite{PDG}, one is able to constrain the range above. The constraint~(\ref{bound_spin2}) sets a stringent bound on the mass of the graviton~\cite{PDG}, and of any massive spin-2 field governed by the field equations~(\ref{GL}).
If the largest known supermassive BHs with $M\simeq 2\times 10^{10} M_\odot$ \cite{McConnell:2011mu,McConnell:2012dh} were confirmed to have nonzero spin, we could get even more stringent bounds.

\section{Conclusions \& Open issues}
\label{CONCLUSIONS}

In this paper we reviewed BH solutions and their stability properties in massive (bi)-gravity. 
This area of research is young and fast developing --- 
the majority of the results here presented have been obtained during the last few years, in the context of the recently discovered dRGT model. 
The main results include: classes of exact solutions featuring solutions of GR (cf. Section \ref{ana_sol}); 
numerical non-GR-like bidiagonal solutions with hair (Section \ref{sec:num}); 
and perturbative (in)stability of BHs in massive (bi)-gravity (Section \ref{PERTURBATIONS}).
However, there are still many questions to be answered and problems to solve. We find it convenient  to close this paper with a (necessarily incomplete and biased) list of outstanding open problems that are, in our opinion, still uncharted territory or not completely understood issues:

\begin{itemize}

\item As shown in~\cite{Volkov:2014ooa}, there is in fact an infinite family of spherically symmetric non-bidiagonal solutions (all related to the Schwarzschild-(A)dS metric through a coordinate change). However, an explicit construction of other solutions, besides the ones studied in Section~\ref{ana_sol}, is still missing. More importantly, a stability analysis of this infinite set of solutions or a study of their global structure (along the lines of~\cite{Blas:2005yk}), could help to decide which solutions are physically relevant.  
 
\item Besides the Kerr solutions discussed here, no other rotating BH solutions have been found so far. Extensions of BHs with massive graviton hair discussed in Sec.~\ref{sec:num} most likely exist. On the other hand the superradiant instability of the Kerr solution could also lead to hairy solutions analogous to the ones found in Ref.~\cite{Herdeiro:2014goa}.

\item Asymptotically de-Sitter hairy solutions have not yet been found. However the instability of the bi-Schwarzschild-de Sitter solution is an indication that they could, in principle, also exist.

\item The classical no-hair theorems of GR do not trivially apply in theories with massive gravitons. Whether massive (bi)-gravity coupled to matter admits hairy BH solutions, besides the extension of standard \emph{(electro)-vacuum} solutions presented here, is unknown. For example, using the linear Fierz-Pauli theory, Ref.~\cite{Deser:2013rxa} suggested that when the matter's energy-momentum tensor has a non-vanishing trace BHs should have hair. However, it is not clear if this is still true for the full nonlinear theory.
   
\item The spherically symmetric instability uncovered in Refs.~\cite{Babichev:2013una,Brito:2013wya} would presumably cause the Schwarzschild spacetime to evolve towards another spherically symmetric solution. However, depending on the parameters of the theory, different types of solutions exist making the end-state of this instability unclear. This in turn makes nonlinear time evolutions of bidiagonal Schwarzschild BHs highly desirable.

\item On another related note, the non-uniqueness of the solutions in massive (bi)-gravity makes it unclear what is the outcome of gravitational collapse. In highly symmetric scenarios the bidiagonal Schwarzschild solutions seem to be the more natural outcome. But it is of course possible that, in some regions of parameter space, Schwarzschild BHs are not the preferred outcome of gravitational collapse or even that BHs do not form in massive gravity. These issues can only be addressed by performing nonlinear collapse simulations.

\item It was argued in Ref.~\cite{Mirbabayi:2013sva} that another type of instability of black holes in massive gravity (with one fixed metric) should exist. 
Such a process would be similar to the discharge of a collapsing charged star, but in this case the black hole's mass would decrease in a finite time (even classically). A complete numerical analysis should be put forward to address this question. 

\item The stability of the hairy solutions discussed in Sec.~\ref{sec:num} remains an open issue. 

\item The absence (or presence) of ghosts in the perturbation's spectrum of BH solutions in massive (bi)-gravity is an open issue. 
Although the construction of the massive interaction terms in dRGT theory and its extension guarantees the absence of the Boulware-Deser ghost, 
the other remaining propagating modes can possibly become ghostly in some backgrounds and must therefore be examined in detail.

\item The linear stability of non-bidiagonal solutions to non-radial perturbations has not been studied. For example, whether the non-bidiagonal Kerr solution discussed in Sec.~\ref{ana_sol} suffers from a superradiant instability is unknown.

\end{itemize}

It is possible that some of the theoretical issues discussed here might help to decide whether massive (bi)-gravity is a viable theory to describe our physical world or not. Thus, 
we hope that this paper will serve as a guide for future developments on this relatively new and exciting topic.

\vspace{0.1cm}
\noindent
{\bf Acknowledgments.}
We would like to thank Paolo Pani and Helvi Witek for the invitation to contribute to the Focus Issue on ``Black holes
and fundamental fields''. We thank Angnis Schmidt-May for useful correspondence. We are indebted to Michael Volkov, Vitor Cardoso and Paolo Pani for useful comments on a preliminary draft of this manuscript. We are also much indebted to Vitor Cardoso, Alessandro Fabbri and Paolo Pani for many and interesting discussions throughout the past few years without whom part of this work would have been impossible.
The work of E.B. was supported in part by the Russian Foundation for Basic Research Grants No. RFBR 13-02-00257, 15-02-05038.
R.B. acknowledges financial support from the FCT-IDPASC program through the Grant No. SFRH/BD/52047/2012, and from the Funda\c c\~ao Calouste Gulbenkian through
the Programa Gulbenkian de Est\' imulo \`a Investiga\c c\~ao Cient\'ifica.
This research was supported in part by the European
Union's FP7 ERC Starting Grant ``The dynamics of black holes: testing
the limits of Einstein's theory'' grant agreement no. DyBHo--256667.

\section*{References}
\bibliographystyle{iopart-num}

\end{document}